\documentstyle[aaspptwo,epsf]{article}
\newcommand{\etal}{{{ et al.}}~}
\newcommand{\eg}{{{e.g.,}}~}
\newcommand{\ie}{{{i.e.,}}~}
\newcommand{\kms}{{{km s$^{-1}$}}~}
\newcommand{\kmsp}{{{km s$^{-1}.$}}~}
\newcommand{\kmsc}{{{km s$^{-1},$}}~}

\slugcomment{To appear in the Astrophysical Journal: Dec. 20 1996}

\begin{document}

\title{The Observational Consequences\\
    of \\
    Merging Clusters of Galaxies}

\vspace{1.5in}
\author{KURT ROETTIGER$^1$, JACK O. BURNS \& CHRIS LOKEN}
\affil{Department of Astronomy \\New Mexico State University 
\\Las Cruces, New Mexico 88003\\
e-mail: kroett@pecos.astro.umd.edu\\cloken@nmsu.edu\\jburns@nmsu.edu}
\vspace{1.0in}
\altaffiltext{1} {Current Address: Department of Astronomy, University of Maryland, College Park, MD 20742}




\begin{abstract}
We present an observational analysis of the numerical simulations
of galaxy cluster mergers presented in Roettiger, Loken \& Burns (1996).
We identify several observational signatures of recent merger activity, and
quantitatively assess the uncertainty introduced into cluster mass 
estimates when invoking the commonly held assumptions of hydrostatic equilibrium, 
virial equilibrium, spherical
symmetry and isothermality. We find that mergers result in multiple X-ray
peaks, long-lived elongation of the X-ray emission as well as isophotal twisting and
centroid shifting to a degree consistent with recent observations. We also
find an enlargement of the X-ray core relative to the dark matter core.
Mergers result in non-isothermal clusters exhibiting observable 
inhomogeneities in the emission-weighted X-ray temperature of several keV on linear scales of less than 0.5 Mpc. The resulting gas dynamics are extremely complex, and we present an
example of what might be observed by a high resolution X-ray spectrograph.
We further speculate that the gas dynamics, via shocks, bulk flows
and turbulence, play an important role in the evolution of cluster galaxies
and associated radio sources, particularly wide-angle tailed (WAT) sources and radio halos. We find that X-ray based
cluster mass estimates made under equilibrium assumptions can be uncertain
by 50\% or more in the first 2 Gyrs after a merger and by up
to 25\% after 2 Gyrs depending on the details of the analysis and
projection effects. Uncertainties can be considerably larger if the
temperature is not well constrained. Similar
uncertainties are observed in the X-ray derived baryon mass fractions.
Virial mass estimates are typically overestimated because the observed
1-dimensional velocity dispersion can be severely contaminated by the infall velocity of the subcluster.

\end{abstract}


\section{INTRODUCTION}

Clusters of galaxies are the largest gravitationally bound objects 
in the Universe. To date, thousands have been identified through
both optical and X-ray surveys. Since they are abundant, luminous, and extended
objects capable of being observed in some detail at cosmologically significant
distances, their value to cosmology is unquestioned. Clusters of galaxies have been
used to constrain the matter density in the Universe, $\Omega$ (\eg Nakamura, Hattori, \& Mineshige 1995), 
the ratio of baryonic to dark matter density in the Universe, $\Omega_b$ (\eg White \& Fabian 1995), and 
even the power spectrum of primordial density fluctuations (\eg Henry \& Arnaud 1991). In each case, 
accurate knowledge of
the gravitational mass in clusters is required. Traditionally, cluster masses have been
estimated from the velocity dispersion and spatial distribution of member galaxies assuming them to be in virial equilibrium (\eg Merrit 1987), 
or from the distribution of X-ray emitting gas assuming it to be in hydrostatic equilibrium (\eg Fabricant, Lecar, \& Gorentein 1980). 
More recently, it has become possible to determine cluster masses from the effects 
of gravitational lensing (\eg Tyson, Valdez, \& Wenk 1990). Since this
is the only direct method of mass determination and is therefore not subject to assumptions
regarding cluster dynamics, it is potentially the most powerful tool. Unfortunately, in
the instances where all three methods have been applied, the
results can often disagree by factors of several thus
raising questions as to the validity of the underlying assumptions in each
method (\eg Loeb \& Mao 1994; Kneib \etal 1995). 

In a hierarchical universe in which large-scale structures form via the accretion of smaller structures, the
greatest threat to equilibrium conditions within a given cluster resides in the potential for a merger with another cluster.
 In order to determine the degree and duration
of non-equilibrium conditions following a major merger event, we have conducted a series of numerical
hydrodynamical/N-body simulations of isolated merging clusters of galaxies. The results of these
simulations have appeared in a series of papers. Paper I 
(Roettiger, Burns \& Loken 1993) presented a brief description 
of the numerical simulations along with several
general results while Roettiger, Loken \& Burns 1996 (Paper II) provided a
detailed description of the numerical simulations and the physical evolution of
clusters during the merger. In this paper, we address the questions: 1) What
are the observable signatures of merging clusters? 2)
How does the violation of the conditions of hydrostatic and virial equilibrium
affect the determination of such cosmologically important parameters as
cluster masses and the baryon to dark matter mass ratio? Understanding the observational signatures
of cluster-subcluster mergers will help us to identify clusters which are not relaxed and may therefore
be subject to considerable uncertainty in their mass estimates.

The numerical simulations and the parameter survey upon which this work is
based are discussed in detail in Paper II. Here, we present only a brief
overview. The numerical simulations employ a hybrid Hydro/N-body code
in which the intracluster medium is represented by the equations of
hydrodynamics, and the dark matter is represented by the N-body particle
distribution. The hydrodynamical component of the code is ZEUS-3D 
(Clarke 1990; Stone \& Norman 1992), an
Eulerian, finite-difference code developed and supported by the Laboratory
for Computational Astrophysics (LCA) at the National Center for
Supercomputing Applications (NCSA). The N-body component of the code
is TREECODE (Hernquist 1987), a hierarchical tree algorithm in which
the force between nearby particles is calculated directly while more
distant particles are grouped so that the force term is calculated via a multipole
expansion. The gravitational potential is defined entirely by the dark matter component. The gas component is chosen to be $<$5\% of the total cluster
mass so as to be dynamically insignificant.  Radiative cooling has not
been included in these simulations. The central cluster densities, nominally 10$^{-3}$ cm$^{-3}$,
were chosen so as to make the effects of cooling negligible.

We begin with two isolated, isothermal, hydrostatic King spheres placed on the computational grid such
that their cores are separated by $\sim$5 Mpc.
They are then allowed to merge under the influence of their mutual gravity.
The idealized initial conditions have several benefits. First, they allow a more
efficient use of the computational volume, and consequently, higher resolution
than is typically found in most cosmological simulations (\eg Bryan \etal 1994).
The resolution of these simulations is 50 kpc or 5 zones across the primary
cluster core radius. Second, they provide a well-defined baseline with which
to compare the subsequent evolution of the clusters after the mergers. And
third, they allow the study of the merger event in the absence of the more
general cluster formation process.

The parameter survey discussed in Paper II was designed to study the
effects of varying the relative cluster-subcluster masses and their
relative gas content. The mass ratios in the survey were 1:1, 2:1, 4:1 and 8:1. 
The variation of the cluster-subcluster mass
ratio implied different values for both the impact velocity and the relative linear dimensions
of the clusters. We also conducted a simulation in which the smaller of 
the two clusters (hereafter, the subcluster) was essentially gas-free. This showed
the relative importance of the hydrodynamical processes (\ie shocks, bulk
flows) in the merger evolution.

In this paper, we address some of the observational
consequences of cluster-subcluster mergers. In \S \ref{samples}, we describe the merger
samples presented in this paper. In \S \ref{xmorph}, we discuss the evolution of the X-ray morphology.
Section \ref{tmorph} contains a discussion of the observable temperature substructure. Section \ref{gdynam} addresses
the observable consequences of the gas dynamics. Section \ref{masses} examines the errors associated with  cluster
mass determinations based on the commonly employed assumptions of spherical symmetry,
hydrostatic and virial equilibrium. Also in \S \ref{masses}, we discuss the implications of uncertain mass estimates
and non-equilibrium conditions on the determination
of the baryon fraction in clusters of galaxies.  Section \ref{summary} is a summary of our results.

\section {THE SAMPLES}
\label{samples}

For the sake of brevity and clarity, we cannot present every observational aspect of every
merger in our parameter survey.  We have therefore chosen two sub-samples having characteristics
representative of the whole.
The first is the evolutionary sample which shows the evolution of the 4:1 mass ratio merger
at six epochs. The second, or deprojection sample, contains several mergers representing
various mass ratios, projection angles and epochs. The chosen epochs are $-0.25$, 0, 0.5, 1.0, 2.0
and 5.0 Gyrs relative to the time of core passage. These epochs are the
same as those presented for all mergers in Paper II thus allowing a direct examination of
the underlying physics at each epoch.

\subsection {The Evolutionary Sample}
\label{esamp}

The evolutionary sample was chosen to depict the evolution of a single mass ratio merger
as viewed in the plane of the sky. We have chosen the 4:1 mass ratio merger as a representative
example of a moderate mass ratio merger. At each epoch, we view it perpendicular to the merger axis
because it is from this angle that the X-ray morphology is most severely disturbed. Observational projection effects
are addressed in the deprojection sample (\S \ref{dsamp}).  To summarize, 
the primary cluster is scaled to 4 $\times$ 10$^{14}$
M$_\odot$ with a velocity dispersion of 660 \kmsp The subcluster is scaled to 1 $\times$ 10$^{14}$
M$_\odot$ with a velocity dispersion of 415 \kmsp The primary cluster gas temperature is 4.6 keV while the subcluster
temperature is 1.8 keV.  The dark matter component of the clusters is represented by 22,500
equal mass particles.  The synthetic X-ray images for this sample can be seen in Figure \ref{evol} (See \S \ref{xmorph}).

\subsection { The Deprojection Sample}
\label{dsamp}

The deprojection sample was chosen to be representative of various
mass ratios, epochs, and merger axis projection angles such that a variety of
realistic and representative observational situations are depicted.
To this end, we have selected six examples for the X-ray deprojection
analysis (\S \ref{masses}). The synthetic X-ray images can be seen in Figure \ref{depro}. The merger 
parameters as well as several observable quantities are listed in Table \ref{tab1.3}. 
Further details are in Table 1 of Paper II. The sample includes two examples with 
obvious X-ray substructure (Mergers 1 and
5, as indicated in Table 1), three examples in which the X-ray morphology appears somewhat 
relaxed although it is known that shocks are present (Mergers 2,3, and 4),
several mergers in which the observed velocity dispersion is contaminated
by the infall velocity to varying degrees (Mergers 1-4), and two examples
in which shocks and bulk gas flows have diminished and contamination
of the velocity dispersion is minimal (Mergers 5 and 6).

\section {THE X-RAY MORPHOLOGY}
\label{xmorph}

In this section, we examine the variety and evolution of the X-ray surface brightness morphologies
exhibited during the merger.  Synthetic X-ray images of the evolutionary and deprojection samples can be seen
in Figures \ref{evol} and \ref{depro}, respectively. The X-ray emission is based on the Raymond-Smith X-ray 
spectrum which includes
both thermal bremsstrahlung and line emission within a ROSAT-like bandpass of 0.5-2.0 keV 
(Raymond \& Smith 1977). Using the 3-dimensional density
and temperature data from the numerical simulations, we have calculated the X-ray volume emissivity in each zone.
The X-ray surface brightness images (Figures \ref{evol} and \ref{depro}) were generated via a line-of-sight (LOS) integration 
through the computational volume assuming the emission to be optically thin. No attempt has been made
to add statistical noise.  The boxy edges in the outer contours are indicative of the reduced
resolution near the edges of the computational volume (see Paper II, \S 3.4).

Figure \ref{evol} shows the evolution of the X-ray surface brightness for the evolutionary sample. Not pictured is the initial cluster which was spherically symmetric. The surface 
brightness has been projected such that the merger axis runs vertically and is in the plane
of the sky. The subcluster entered from the top of the panel and proceeded downward. 
At the top of each panel, we indicate the merger epoch relative to the time of core passage.
The contour levels are logarithmic and constant from one epoch to the next.

In Figure \ref{evol}, we see a variety of X-ray morphologies. Several epochs, both before and after core passage, exhibit
multiple concentrations of X-ray emission (Figures \ref{evol}a,c,d,e). In this merger, the subcluster penetrated
the core of the primary and exits on the downstream side resulting in a secondary maximum below the
X-ray core. All epochs after core passage exhibit a general elongation of the X-ray 
morphology parallel to the merger axis. 
As was shown in \S 5.5 of Paper II, the elongation of the X-ray (\ie gas density) distribution is often 
more extreme than that of the underlying dark matter distribution.  We also note that the elongation of the X-ray morphology is not limited to being parallel to the merger axis. Figures \ref{evol}b and d and to a lesser extent, e, show
distinct elongation of the central region perpendicular to the merger axis. Although not apparent
in Figure \ref{evol}c because of the choice of contours, both epochs corresponding to Figures \ref{evol}c and d show ``bars"
of X-ray emission near the cluster core. An equivalent feature is not apparent in the dark matter distribution indicating
a severe violation of hydrostatic equilibrium.  The elongation of the gas distribution perpendicular to the merger
axis is due to a shock and adiabatic compression generated by the infalling subcluster. This morphology is quite similar to that of A754 (Henry \& Briel
1995), and Henriksen (1993) has suggested that A754 is the result of a recent merger between two unequal mass clusters.
   
Finally, even after 5 Gyrs (Figure \ref{evol}f), we still see the effects of
the merger in the X-ray morphology. On the downstream side of the core there is a flattening of the X-ray contours as a result of residual infall from the subcluster's return trip through the primary core.  Also notable in Figure \ref{evol}f 
is the much expanded X-ray distribution (compare with Figure \ref{evol}a). 
One important consequence of the merger is an increase in the cluster X-ray core radius, often by a factor of
2, relative to the dark matter core. The expansion of the gas core is 
largely thermally supported as a result of
 shock heating. However, in the early stages of the merger dynamical support (\ie bulk flows and turbulence) is also important. At 0.5 and 1.0 Gyr, the
mean dynamical pressure ($\rho v^2$) within the core is nearly 30\% of
the mean thermal pressure ($nkT$) within the same region. At 2.0 and 5.0 Gyrs
the mean dynamical pressure has dropped to less than $\sim$15\% of the mean thermal
pressure.  Figure \ref{prof} demonstrates both the absolute expansion of the X-ray core
as well as the expansion of the X-ray core relative to the dark matter core.

The significance
of the relative core sizes is two-fold. First, recent gravitational lensing experiments
find that dark matter cores are significantly smaller than the X-ray cores 
(\eg Miralda-Escud\'e \& Babul 1995) leading to lensing-based mass estimates 2-3 times larger than those
derived from the X-ray emission. Second, the standard X-ray deprojection technique (Arnaud 1988)
requires an estimate of the dark matter or gravitational core radius. Since the dark matter is unobservable, the
core radius estimate comes from the X-ray surface brightness image. The consequences of this
discrepancy are discussed in \S \ref{masses}.

Figure \ref{depro} contains the synthetic X-ray surface brightness images for the deprojection sample.  Again, the epochs depicted are the same as in Figure \ref{evol} only now the
mass ratios and merger axis projection angles differ. The X-ray morphologies shown in Figure \ref{depro} are quite
similar to those in Figure \ref{evol}. However, the projection can often serve to soften features and hide evidence of a recent merger. This
is important since the X-ray emission is often used as an indicator of the evolutionary state of the cluster 
(\eg Mohr \etal 1995).

One of the best examples of how projection effects can hide substructure is Figure \ref{depro}d in which a 2:1 mass ratio merger is projected 45$^\circ$ from the plane of the
sky 1 Gyr after core passage. If viewed in the plane of the sky, one would witness a considerably more prominent
extension below the cluster core as well as extensions perpendicular to the
merger axis on either side of the core (\eg Figure \ref{evol}d). These features are almost completely hidden 
in this projection. In order to obtain perspective on 
the projection effects, one can compare the projected X-ray morphologies in 
Figure \ref{depro} with the gas density contours for the corresponding mass 
ratio and epoch in Figures 6-9 of Paper II. 

In the absence of obvious substructure, we can look for more subtle evidence of 
merger activity such as isophotal twisting or centroid shifting. The results of this test are shown in Figure \ref{ellipses} where
we have used the IRAF routine ELLIPSE to fit elliptical isophotes to the synthetic X-ray distribution in Figure \ref{depro}d.
In this case, the ellipticity varies dramatically with
radius from near 0 at large radii to 0.42 near the core. Although there does appear to be some minor isophotal
twisting, all position angles of all ellipses are consistent at the 2-$\sigma$ level with alignment along the merger axis. We also see statistically significant centroid shifts of $\sim$150 kpc. Nearly the entire 
magnitude of the centroid shift occurs along the merger axis. The magnitude of the maximum centroid shift in this
example is only slightly larger than the largest reported by Mohr \etal (1995) for their sample of clusters observed
by the {\it Einstein} satellite.

\section {THE GAS TEMPERATURE MORPHOLOGY}
\label{tmorph}

The gas temperatures within the ICM can be an important diagnostic of
the dynamical state of the cluster. Substructure, shocks and adiabatic cooling can result in a very patchy and distinctly non-isothermal gas distribution. This is in stark contrast to the commonly held notion of an isothermal
cluster. Because of limited spatial and spectral resolution in previous X-ray observations, it is only recently that the level of inhomogeneity
in the temperature morphology has become fully appreciated (\eg Briel \& Henry 1994; Arnaud \etal 1994). 
The situation, however, is rapidly changing since {\it ASCA}, and, eventually {\it AXAF}, promise to
supply even more accurate and detailed temperature maps in the near future. For this reason, we focus here on the
observable temperature morphology that results from cluster-subcluster mergers.

Figures \ref{temp1} and \ref{temp} show the observable temperature maps (solid contours, labeled with the temperature in keV) 
 overlaid onto the X-ray surface brightness (dotted contours) for the evolutionary (\S \ref{esamp}) and deprojection
{\S \ref{dsamp}) samples,
respectively.  The temperature maps are generated by calculating
a mean emissivity-weighted temperature along the LOS through the cluster. The temperature in each zone is weighted by 
the volume emissivity of that zone within a  ROSAT-like bandpass of 0.5-2.0 keV. The true cluster gas temperatures 
can be seen in Figure 12-15 of Paper II. Comparing these figures shows that projection effects can severely alter the temperature morphology. In general, the projection
and LOS integration tend to mitigate the extreme temperatures seen in the various mergers.

As an example of  projection effects, Figure \ref{temp}a shows the 2:1 mass ratio merger shortly
before core passage and projected at 45$^\circ$ to the plane of the sky. This is a similar situation to the A2256
model presented in Roettiger, Burns, \& Pinkney (1995) although the scaling is different. The true
temperature in a central slice along the merger axis is shown
in Figure \ref{texample}. Close examination shows the relatively
cool subcluster (top) impinging on the somewhat warmer primary cluster (bottom). A shock has formed at the cluster-subcluster
interface and is visible as a thin layer of very hot gas perpendicular to the merger axis. This region is actually
a broad conical structure whose axis of symmetry coincides with the merger axis. The hottest gas is located at the apex of the cone.  In Figure \ref{temp}a,
the conical region of hot gas appears as two ``hot lobes"  located on either side of the
merger axis (vertical line through the X-ray maxima). Because of the 45$^\circ$ projection, the conical structure is seen more
head-on and is therefore broader than the more linear feature seen in Figure \ref{texample}. It takes on the lobe structure
because the temperature in the central region is contaminated by cooler subcluster gas projected along the
same LOS.  Figure \ref{temp}b exhibits the expected morphology when the conical shock front is viewed in the plane
of the sky.

A variety of complex temperature morphologies are exhibited throughout the rest of the deprojection sample.
In Figure \ref{temp}c, there are four distinct temperature regions located along the merger axis. 
Starting at the top and moving
downward, there is a region contained within 7.5 keV. This region is actually a local minimum ($<$7.5 keV)
 and is identified as the relatively cool wake of the infalling subcluster.
There is a hotspot ($>$16 keV) located at the X-ray core. Below the X-ray core, 
there is another cool region that we identify as the emerging subcluster. Below the subcluster there
is a large arc of shock-heated gas, $>$ 11 keV.
A similar morphology is seen in Figure \ref{temp}d though in this case, the hotspot, once located near the core, 
is seen to have propagated upstream. This is the result of a shock that forms when gas that is expelled 
from the core by the rapidly
varying gravitational potential impinges on residual infall from the subcluster.
To summarize, we find regions of gas varying by several keV that are separated on linear scales of several 
hundred kiloparsecs.

As one might expect, the most complex temperature structures will be those viewed perpendicular to the merger
axis and within 2 Gyrs after core passage. Figure \ref{temp}e reveals a hot core which, to varying degrees, results
from shock heating and a continuous stirring of the gas core by the varying gravitational
potential. It also shows a significantly cooler region downstream from the merger. This appears to be analogous
to a cool region noted by Schindler (1996) which was identified as a rarefaction wave following the initial bow shock.
Although this is a possible explanation, there is evidence in our simulations that this is actually gas from 
the subcluster that has passed through the core. 

As described in \S 5.2.1 in Paper II, we believe that the subcluster gas
is protected from ram pressure stripping by the bow shock that forms on the
subcluster's leading edge shortly before core passage. The effect
of the bow shock is to reduce the ram pressure by reducing and
redirecting the flow relative to the subcluster core. We tested this interpretation by running 
the 4:1 mass ratio simulation
with a passive-scalar, (a dynamically inert quantity used to trace the various gas
components in numerical simulations, see \S5.2, Paper II), attached to the subcluster and found that the passive-scalar did penetrate the core.

Since the gravitational potential of the subcluster is largely disrupted
after core passage, it no longer confines the subcluster gas, and consequently, the subcluster gas expands and cools 
adiabatically. 

Finally, Figure \ref{temp}f is viewed perpendicular to the merger axis but
at this stage, 5 Gyrs after core passage, we find a relatively relaxed configuration in which the shocks
have dissipated and the subcluster is thoroughly mixed. The centrally peaked temperature distribution is
maintained by a continual mixing of the gas by the varying gravitational potential minimum. 

The results presented here are quantitatively very similar to recent temperature maps generated from both
{\it ROSAT} and {\it ASCA} data. We have made a detailed comparison between these models and the 
{\it ROSAT} temperature map of A2256 presented by Briel \& Henry (1994) in
Roettiger \etal (1995). 
Henry \& Briel (1995) present a temperature map of A754 which shows inhomogeneities of similar scale.  Initial
{\it ASCA} results have been published for A2163 (Markevitch \etal 1994) and Perseus (Arnaud \etal 1994). Both
show widely varying temperatures ($\Delta$T $\sim$ 5 keV) on scales $<$0.5 Mpc. In addition to the
temperature morphology, there is other evidence that 
A2163 is in the process of a merger.  Miralda-Escud\'{e} \& Babul (1995) find evidence from gravitational lensing
that the gas distribution is not in hydrostatic equilibrium with the gravitational potential, and Herbig \&
Birkenshaw (1994) have identified a radio halo associated with A2163. Cluster radio halos are a Mpc-scale
phenomenon which has been linked to cluster-subcluster mergers (Jaffe 1992; Tribble 1993; Burns \etal 1994, see \ref{halo}).

\section {THE GAS DYNAMICS}
\label{gdynam}

There are several important observational consequences of the
gas dynamics induced by the cluster-subcluster merger. As we have seen,  mergers result
in shocks, bulk flows, and large eddies which appear to decay into a random velocity
field on timescales of $\sim$2 Gyrs. The indirect effects of the gas dynamics on 
the accuracy of equilibrium-based mass estimates will be
addressed in \S \ref{masses}. In this section, we discuss the more directly observable
consequences.

\subsection {Spectroscopic Gas Velocities}
\label{gvel}

Currently, it is not possible to directly measure the dynamics of the
X-ray emitting ICM. However, this will change with the planned launch 
of {\it ASTRO-E} in 2000. So, with the launch date rapidly approaching, it is now time to
make predictions as to the observable velocity field that may result from 
a cluster-subcluster merger. In this section, we use the
results from  the numerical simulation to determine quantitatively 
the level and structure of the observable gas dynamics as they might be viewed 
by a high spectral/spatial resolution X-ray telescope. 

As with other astronomical
spectroscopy, X-ray spectroscopy will only allow the determination of
LOS velocities and, assuming optically-thin emission, the X-ray spectrum
at any point will be a composite of all emitting regions along the
LOS. We have not, at this time, attempted a detailed model of the cluster 
X-ray spectrum nor have we taken into account the specific characteristics of {\it ASTRO-E}. We have, however, 
attempted to model an ``observable" velocity map based on the projected LOS gas velocities weighted by the corresponding Raymond-Smith X-ray emissivity.  Since the X-ray spectral lines are recombination lines, the line emissivity, 
like that of the thermal bremsstrahlung, is expected to scale as $n^2$ (\eg Sarazin 1986).

As an example of an observable velocity map, Figure \ref{velocity}a is a subregion of Figure \ref{depro}d and contains the mean LOS emission-weighted gas velocity
for Merger 4 of the deprojection sample (see Table \ref{tab1.3}). Similarly, Figure \ref{velocity}b contains
the emission-weighted gas velocity dispersion along the same projected LOS. All velocities are
in the rest frame of the dark matter core of the primary cluster. As a spatial reference, we have included in Figure \ref{velocity}
contours of the X-ray surface brightness (dotted line).
 
Merger 4 is a 2:1 mass ratio merger at 1 Gyr after core passage.  
The merger axis is projected 45$^\circ$ out of the plane of the sky such that the subcluster is 
falling toward the observer (negative velocity). The region
of relatively large negative velocity, $<-450$ \kmsc located below the X-ray peak is subcluster
gas that has passed through the core of the primary. As with all other figures, the subcluster
originated at the top of the panel and proceeds downward.  The effect of the merger axis projection 
and the LOS integration is to soften the velocity extremes. The true velocity structure, in a
plane parallel to the merger axis, can be seen in Figure 25d of Paper II. The peak velocity in that
figure is in excess of 1000 \kmsp In Figure \ref{velocity}a, the absolute maximum is
$\sim$550 \kmsp  The effect of the emission-weighting is to bias the LOS integration toward the 
denser and therefore more dynamically (and observationally) significant central regions of the cluster. 

The velocity dispersion map is also emissivity-weighted. We find the greatest velocity 
dispersions to be near the edges of the advancing subcluster (two peaks
at 350 \kms below the X-ray core) where gas is being sheared away and rapidly decelerated.  A large 
dispersion is
also apparent in the region above the X-ray core where there is still some residual 
infall that is being decelerated as it collides with gas expelled from the core.  It 
should be kept in mind that this is only one
example and is subject to strong projection effects. It is, therefore, meant only to serve as an illustrative
example.

\subsection {Gas Dynamics and Cluster Radio Sources}

\subsubsection { The WAT Problem}
\label{wat}

Generally, radio tails are found to be underpressurized with respect to the ICM and are therefore
subject to the dynamical influences of the thermal X-ray gas (Feretti, Perola, \& Fanti 1992). It has long
been suggested that the narrow-angle tailed (NAT) radio source morphology results when the radio tails
are swept back by ram pressure as the host galaxy moves at high speed through the ICM (Miley \etal 1972). Although this
explanation works well for NATs, it does not work well for wide-angle tailed radio sources or WATs (\eg Eilek \etal 1984).
WATs are C-shaped radio sources consisting of two, straight and narrow jets that flare into broad
tails which, like NATs,  extend in a common direction. The jets typically extend 10-100 kpc while the
total extent of the source, including tails, can reach 1 Mpc. Unlike NATs, however, WATs are associated
with the central dominant galaxies in clusters and are therefore presumed to be at rest with
respect to the cluster gravitational potential and, assuming hydrostatic equilibrium, the ICM. 
What then is bending the WAT? (see also Eilek \etal 1984; O'Donoghue, Eilek \& Owen 1993)

We have suggested that bulk flows generated during mergers may be responsible
for the bending of the WAT tails (Roettiger \etal 1993; Pinkney, Burns \& Hill 1994; Loken \etal 1995). The numerical simulations
presented here allow us to add quantitative support to this idea.
 Previous analysis of the radio data indicates that ICM flow
velocities on the order of 1000 \kms are required, assuming typical central gas densities
($\sim$10$^{-3}$ cm$^{-3}$),
to provide the necessary ram pressure to bend the jets
(\eg Burns \etal 1986). The linear scale of WATs, ranging from 100 kpc to 1 Mpc, gives some indication as to
the required linear dimensions of the bulk flows. These conditions are met for extended 
periods during the evolution of a merger.

We find peak gas velocities well in excess of 1000 \kms at various stages of the merger evolution. 
Figure \ref{vbulk1} shows that the peak gas velocities generally do not
decay below 1000 \kms for nearly 2 Gyrs after core passage. This is much longer
than the estimated ages of the radio sources, 10$^7$-10$^8$ years.  Figure \ref{vbulk2} shows the level of coherence
to the flow as a function of time. Typically the bulk flow through the cluster core is $>$200 kpc wide and
can be $\sim$1 Mpc wide. The coherence of the flow diminishes on time scales of several Gyrs. Finally, we find
bulk flows in both directions along the merger axis at various times during the evolution of the merger.

If the merger hypothesis is true, one would expect WAT clusters to exhibit other merger signatures
such as an elongation of the X-ray morphology in the same direction as the radio tails, an absence of cooling
flows, velocity substructure, etc. All of these are indeed observed. Velocity substructure in WAT clusters
is discussed in  Pinkney (1995). {\it ROSAT} observations showing elongations of X-ray emission along
the direction of the WAT tails are presented in G\'omez \etal (1996) and in
Pinkney \etal (1993, 1994). Numerically, this hypothesis has been addressed more directly by Loken \etal (1995)
with numerical simulations of radio jets/tails within the velocity field of a merging cluster like that in 
Figure 24c of
 Paper II.

\subsubsection { Radio Halos}
\label{halo}

Radio halos are large ($\sim$ 1 Mpc), amorphous cluster sources of uncertain origin and generally steep 
radio spectra (see Hanisch 1982). They are extremely rare sources that appear to be 
associated only with very rich clusters that have
undergone recent mergers. Some of the best examples include Coma, A2255 and A2256. Each of these clusters
has been reported to have undergone a recent merger based on a variety of properties including X-ray morphology, 
galaxy velocity distributions, temperature morphology and the absence of cooling cores (Burns \etal 1994, Burns \etal 1995, Roettiger
\etal 1995, Briel \etal 1991). Therefore, as has been
suggested by various authors (\eg Tribble 1993; Burns \etal 1994), it may be reasonable to expect that cluster
halos are also related to recent merger activity. 

The difficulty in explaining radio halos arises from the
combination of their large size, up to 1 Mpc, and the short synchrotron lifetime of the
relativistic electrons. The expected diffusion velocity of the electron population is on the order of the
Alfven speed ($\sim$100 \kms) making it difficult for the electrons to diffuse over a Mpc-scale region within
their radiative lifetime.
Therefore, one needs a mechanism by which the relativistic electron population can be transported over
large distances in a short time, or a mechanism by which the local electron population is reaccelerated
and local magnetic fields are amplified over an extended region.  The cluster-subcluster merger
can potentially supply both mechanisms.

The first mechanism has been used to explain the halos in A2256 (R\"ottgering \etal 1994) and Coma (Giovannini \etal 1993).
R\"ottgering \etal (1994) contend that the halo source in A2256 results from the distortion of several pre-existing head-tail
sources by the turbulence and shear generated by the infalling subcluster. They calculate that the
relativistic electrons would need to be transported at 1500-3000 \kms in order to account for the extent of
the source. In the A2256 merger model presented in Roettiger \etal (1995), the infalling subcluster has a velocity
of nearly 3000 \kms at the current epoch. 

The second mechanism has been applied to the Coma cluster radio halo (Burns \etal 1994).
Here, the radio halo results from the global disruption caused by the merger. During the merger, shocks and
turbulence create sites for in-situ particle reacceleration and amplification of the magnetic field over large scales
(Eilek \& Hughes, 1990; Ruzmaikin \etal 1989). A simple calculation of the total kinetic power imparted by the
merger (\eg Burns \etal 1995) shows that the merger can potentially supply many orders of magnitude
more energy than is necessary to power the radio source, 10$^{46}$ vs. 10$^{41}$ ergs s$^{-1}$ for Coma and A2255 (Burns \etal 1994; Burns \etal 1995). Although the details of this mechanism are not fully understood, if we accept this
premise, the lifetime of the halos will be governed not by the radiative lifetime of the electron population but rather by
the time scale for the dissipation of the shocks and decay of the turbulence.  Looking at the bulk velocity
properties presented in Figures \ref{vbulk1} and \ref{vbulk3}, we find that the e-folding time for both the peak and mean gas velocities associated with the central 1 Mpc of the primary cluster is greater than 1 Gyr. The shocks dissipate on similar timescales. This
opens the possibility that radio halos may live considerably longer than the 10$^8$ yrs suggested by Tribble (1993).
As Tribble noted, one benefit of the short lifetime is that it serves to explain the rarity of radio halos.
However, their rarity may be attributable to the relative rarity of suitable hosts. To date radio halos have only been identified in 
very rich clusters (richness class $>$2). Finally, if both proposed mechanisms are responsible for the generation of halos, it may be that each is important during different stages of the merger evolution. A2256 appears to be in the very early stages of its merger (Roettiger \etal 1995) while Coma is in the relatively late stages of its merger (Burns \etal 1994).

\subsection {Gas Dynamics and Galaxy Evolution}
\label{galaxy3}

The galaxies found in clusters of galaxies are systematically different from those 
in the field (see Oemler 1992 for a review).
Cluster galaxies are typically redder and contain less gas than isolated field systems. They are also
more likely to be ellipticals than spirals, particularly in the cores of rich clusters.
This has led many to argue that the cluster environment plays an important role in the evolution of
individual galaxies (\eg Spitzer \& Baade 1951; Gunn \& Gott 1972). 

One mechanism by which the environment may influence galaxy evolution is
via the galaxy-ICM interaction.  Dressler \& Gunn (1983) suggested that field galaxies fall into
the cluster environment where they encounter shocks in the ICM. As a result of the shock encounter,
the galaxies experience an increase in ram pressure which may induce a burst of star formation and/or
strip the galaxy of its interstellar medium (ISM). Numerical
simulations of cluster formation show that during the formation process a spherical accretion shock 
forms near the core of the cluster and propagates outward effectively forming an outer boundary to 
the cluster (Evrard 1990). 
Evrard (1991) showed that field galaxies undergoing a burst of star formation upon interacting with
the accretion shock may explain many of the properties of Butcher-Oemler galaxies observed at moderate redshifts 
(Butcher \& Oemler 1978). Edge \etal (1990) suggested that cluster mergers may also contribute to 
the Butcher-Oemler effect since the apparent galaxy evolution occurs during a period of significant
evolution in the cluster X-ray luminosity function (z $<$ 0.5).

A bow shock forms on the leading edge of the subcluster as it approaches
the core of the primary. This shock serves to protect the subcluster from being stripped of its
own ICM by creating a boundary layer that effectively reduces the ram pressure by decreasing
the flow velocity and redirecting it around the core of the subcluster (see Figure 10 of Paper II). In doing so, the shock
will also serve to protect the relatively gas-rich subcluster galaxies. The protection
fails, however, as the cores become coincident. As the shock approaches the core of the primary, it encounters a
rapidly increasing density profile and decelerates 
with respect to the subcluster dark matter and presumably the subcluster galaxy population.
This allows the galaxies to pass through the shock initiating a burst of star formation
followed by a rapid stripping of their ISM which effectively truncates the starburst. It is also possible
that the subcluster galaxies are already actively forming
stars and that interaction with the bow shock simply truncates the existing star formation.  In either case, 
it is this sort of truncated star formation that is believed
to be the mechanism by which post-starburst galaxies are formed (Dressler \& Gunn 1983; Newberry, Boroson \& Kirshner, 1990).

Using our numerical simulations and assuming that the dark matter particle dynamics are representative
of galaxy dynamics, we can track the ram pressure history of individual particles as a function of
their initial cluster membership. We find that the subcluster particles, on average, do experience an
impulsive burst of ram pressure as the cluster cores merge (see Figure \ref{ram}). In fact, the median ram pressure
experienced by subcluster particles increases briefly by fully two orders of magnitude over the initial value.
We have already used this mechanism to explain the post-starburst galaxies observed in the bridge between Coma 
and the NGC 4839 galaxy group (Burns \etal 1994;
Caldwell \etal 1993). These galaxies are dispersed in a manner consistent with the subcluster N-body
particles in the merger simulation. That is, they are distributed along an extension in the X-ray emission
connecting Coma and the NGC 4839 group. For the higher mass ratio mergers, as we believe Coma has recently
undergone, the induced ram pressure is very impulsive. We were therefore able to use the short duration
of the burst in conjunction with starburst evolution models to constrain the age of the merger.

In more general terms, our simulations show that cluster-subcluster mergers can potentially introduce expansive shocks 
into the cluster environment. This is quite apparent from Figure \ref{mach}.
In this Mach number contour plot, the shocks are indicated by strong gradients or closely spaced
contours. Generally, we see two shocks form as a result of the merger. They often appear as two opposing
C-shaped structures in the Mach number figures. The first shock to form is the initial bow
shock described above. This shock propagates downstream while expanding and eventually
dissipating. A second shock, as described in \S \ref{tmorph}, forms after core passage and propagates upstream.
As is apparent in Figure \ref{mach}, these structures can be of Mpc-scale and can sweep
through  Mpc-scales before completely dissipating and thus may influence the evolution of many 
subcluster and primary cluster galaxies.

\section {EQUILIBRIUM MASS ESTIMATES}
\label{masses}

The various methods used by observers to estimate the masses of galaxy clusters often require the assumptions of
hydrostatic equilibrium, virial equilibrium, and spherical symmetry.
In this section, we attempt to reproduce observational conditions and test these methods on the simulated
clusters during various stages of the merger evolution.  Under equilibrium conditions,
the total mass enclosed within a radius, {\it r}, is
given by Fabricant \etal (1980) as,
\begin{equation}
\label{eqmass4}
M(r)={\frac{kT(r)r}{G \mu m_p}}
\left[ -{\frac{dln\rho(r)}{dlnr}}-{\frac{dlnT(r)}{dlnr}} \right],
\end{equation}
where $\rho(r)$ is the gas density profile, $T(r)$ is the gas temperature
profile, {\it G} is the gravitational constant, {\it k} is Boltzmann's constant, $m_p$ is the proton
mass, and $\mu$ is the mean molecular weight
which we have taken to be 0.6.  As is apparent from eq. \ref{eqmass4}, the total mass enclosed within a given radius is
dependent on the local temperature and density gradients at that radius. For this reason, the mass profiles
presented in the following sections may, on occasion, decrease with radius. The ``true" gas density
and temperature profiles can be obtained directly from the simulated data. Observationally, this information
is often not supplied directly by the data. However, both the gas density and temperature
profiles can be obtained through use of the standard X-ray deprojection technique. 

The X-ray deprojection technique, which is described in more detail by Arnaud (1988), uses
the assumption of spherical symmetry to obtain the X-ray volume emissivity from
the observed X-ray surface brightness. It then
uses the assumptions of quasi-hydrostatic equilibrium within spherical shells
in addition to virial equilibrium and isothermality of the dark
matter distribution to
obtain the desired gas density and temperature profiles. We have modified the deprojection technique
to allow for non-isothermality of the dark matter distribution by including a velocity dispersion profile. 
Although the velocity dispersion profile is easy to obtain from the numerical simulations, it is very
difficult to come by observationally. Since we know that our simulated clusters have non-isothermal
dark matter distributions, we assume knowledge of the true velocity dispersion profile in all the following
analysis. Neglecting this correction is equivalent to underestimating the velocity dispersion and can result in a systematic under estimation of the mass in the
central regions of even relaxed clusters (see \S \ref{initmass}).

Owing to the strong dependence of thermal
bremsstrahlung and line emission on the gas density ($\propto n^2$), the density profile
is well constrained by the X-ray emission. The temperature dependence
is considerably weaker ($\sim T^{1/2})$, and consequently, the temperature
profiles are considerably less well constrained. Examination of eq. \ref{eqmass4}
reveals how unfortunate this is since the mass is directly proportional
to the temperature. The temperature uncertainties are less important 
with regards to the temperature gradient which is usually small compared to the gas
density gradient. Fortunately, the gas temperatures can also be obtained through spectroscopic
methods. Unfortunately, high spatial resolution has not coincided with
high spectral resolution.  Therefore, at least until very recently,
cluster temperatures have been  emission-weighted mean temperatures forcing
the assumption of isothermality. As mentioned before, {\it ASCA} and {\it AXAF}
should remedy this situation.

With this in mind, we now estimate the masses of the simulated clusters
using a variety of methods. As a baseline for judging uncertainties in the
mass estimates, we begin by estimating the mass of our initial, spherically symmetric,
hydrostatic primary cluster. We then use the known gas density and 
temperature profiles and eq. \ref{eqmass4} to estimate the cluster masses in the
evolutionary sample (\S \ref{esamp}). Since we are using the true profiles and there are no projection
effects, this analysis represents a best case scenario in which only the assumptions
of hydrostatic equilibrium and spherical symmetry are being tested. We then analyze the
deprojection sample (\S \ref{dsamp}) within the constraints of true cluster observations.  That is, we
deproject the synthetic X-ray images obtaining gas density and temperature profiles
which are then applied to eq. \ref{eqmass4}.  We also analyze the deprojection sample substituting a
mean spectroscopically derived isothermal temperature for the deprojected temperature
profile. Then, having determined the cluster gravitational and gas masses via deprojection,
we examine uncertainties in the derived baryon fraction. Finally, we examine the uncertainties
in virial mass estimates. 

\subsection { The Hydrostatic Spherical Cluster}
\label{initmass}

Figure \ref{staticmass} shows the true mass profile for the primary cluster in the
4:1 mass ratio merger. Also included in Figure \ref{staticmass}
are the mass profiles derived from applying eq. \ref{eqmass4} to: 1) the true gas density and temperature
profiles and 2) the deprojected profiles out to a radius of 2 Mpc.
The mass profile derived from the true profiles is quite accurate. The
RMS variations for $r>$0.25 Mpc are only $\sim$2\%.  The mass profile 
derived from the deprojected profiles are systematically low by $\sim$6\%
with RMS fluctuations of $\sim$3\%. In general, underestimates of the mass
result from underestimates of the temperature. As discussed below, systematic
errors in the deprojected temperature profile are difficult to avoid.

The X-ray deprojection method requires as inputs, in addition to the
X-ray surface brightness profile, a velocity dispersion, a core radius
and an initial guess of the temperature (or pressure) at the outer edge
of the profile. The velocity dispersion is obtained from spectroscopic
redshifts of member galaxies. The core radius, $r_c$, results from fitting
a $\beta$ model (Sarazin 1986),
\begin{equation}
\label{betamod}
S_{X}(r)=A \left [ 1 + \left ( \frac {r} {r_c} \right )^{2} \right ]^{-3\beta + 0.5}
\end{equation}
to the X-ray surface brightness profile, $S_{X}$. The temperature is
often obtained by X-ray spectroscopic methods (\eg David \etal 1993). All three will affect the resulting gas density and temperature
profiles, and ultimately the mass profile to varying degrees. In terms
of the mass profile, the estimate of the outer temperature
is not significant as long as the profile extends beyond the bulk of 
the cluster mass. This is equivalent to saying that the choice of the
outer temperature does not strongly influence the inner temperatures since it
is an iterative solution.
Of course, if the observed X-ray surface brightness profile extends $<$1 Mpc, as is often the
case, then the choice of the outer temperature must be a good one. At
2 Mpc, the deprojection-based mass estimate is not sensitive to the choice of
the outer temperature at least for these relatively concentrated
clusters.

The choice of velocity dispersion and core radius are more important
because they define the gravitational potential within which the gas
is supposedly in hydrostatic equilibrium.  Figure \ref{syserr}a shows how errors
in both the velocity dispersion and the core radius are reflected in the
deprojected temperature profile. A 20\% error in the core radius
(only 1 zone in the simulation), results in a nearly 30\% error in the
central temperature. Fortunately, the problem is well localized in the
core of the cluster and the error in the mass estimate, at least in this
example, is only $\sim$10\%. This is not to say that the mass error will
not be large when the uncertainty in the core radius is large. At this time, it is important to note that
the core radius one wishes to use in the deprojection is
the dark matter (or gravitational) core radius. However, we can only measure the X-ray core
which, as shown by these simulations (\S \ref{xmorph}) and
various gravitational lensing experiments (Wu 1994; Miralde-Escud\'e \& Babul 1995), may be two or more times larger
than the dark matter core.

The temperature is even more sensitive to the velocity
dispersion. A 20\% error in the velocity dispersion can cause a
nearly 50\% error in the central temperature, and the large errors are not localized
to the core region as before. The corresponding errors in the mass estimates approach 40\%.
Errors in the velocity dispersion can easily exceed the conservative numbers used in this example.
Typical uncertainties are $\sim$10-25\% (\eg Girardi \etal 1993), but these are formal errors and
do include systematic errors introduced by anisotropy in the velocity distributions. Mergers produce bulk streaming
and sustained anisotropic velocity dispersions both of which result in contamination of the velocity 
distribution and potential projection effects. There 
is also a potential bias between galaxy and dark matter velocity dispersions resulting from
dynamical friction effects. The velocity dispersion required by the deprojection
is actually the dark matter velocity dispersion which, according to numerical simulations (Frenk \etal 1995), may exceed
the observable galaxy velocity dispersion by as much as 20-30\%.

Figure \ref{syserr}b shows that uncertainties in the core radius and velocity
dispersion do not have a significant effect on the deprojected gas density profiles.

\subsection { Hydrostatic Mass Estimates: True Profiles}
\label{truemass}

In this section, we estimate the cluster masses of the evolutionary sample
 assuming hydrostatic equilibrium and
spherical symmetry. We have azimuthally averaged the simulated 3-dimensional density and
temperature data in spherical shells of $dr$=50 kpc. The resulting
profiles have been applied to eq. \ref{eqmass4}, and the results can be seen in 
Figure \ref{eqmass1}.  

Figure \ref{eqmass1} shows the true mass profile (dashed)  overlaid
with the mass profile derived from eq. \ref{eqmass4} (solid) 
out to a radius of 2 Mpc.  We find localized regions where the mass
can be overestimated by more than a factor of 2. These regions are
associated with shocks in the ICM.  As time passes,
the shocks propagate away from the cluster core and dissipate. The resulting
discrepancy between the true and estimated mass profiles likewise decreases and
appears at larger radii. Whereas the mass is overestimated in the region of
the shock, the mass is underestimated in the region behind the shock
by nearly 40\%.  The greatest errors in the mass determination occur
shortly before and within 2 Gyrs after core passage. This point is
demonstrated graphically in Figure \ref{eqmerr1}.

Figure \ref{eqmerr1} shows the evolution of the mean {\it absolute} error in the hydrostatic equilibrium-based mass estimate.
We have averaged the absolute fractional discrepancy between the estimated and true
mass profiles for $r>$0.25 Mpc. Also included is the RMS of the errors over the same range in radius. Peak-to-peak errors are approximately 
three times the RMS errors. Again,
the errors are most significant, 25\% mean and 40\% RMS, in the first 2 Gyrs after core passage.
This is not surprising since it is during  the first 2 Gyrs that the gas
dynamics play the greatest role and when the conditions for hydrostatic
equilibrium are most severely violated.  It is also not surprising then that the shape of Figure \ref{eqmerr1} mimics 
that of Figure \ref{vbulk3} which shows the evolution of the mean gas dynamics for
each of the simulations. At later epochs, the shocks and bulk flows have somewhat dissipated
 and both the mean and RMS errors are of order 10\%.
There is a slight trend toward systematically low mass estimates during the later stages of the merger.

\subsection {Hydrostatic Mass Estimates: Deprojected Profiles}
\label{depromass}

In this section, we estimate the cluster masses of the deprojection sample based on the deprojected
gas density and temperature profiles.  As stated in \S \ref{initmass}, the deprojection requires an estimate of the temperature at the outer edge of the profile, the velocity dispersion of the dark matter, and the core radius.
The outer temperature was chosen to be the true temperature obtained
from the simulation data.
The velocity dispersion is a deprojected 1-dimensional dispersion  profile along the ``observer's"
LOS at the appropriate projection angle.  The core radius was obtained by fitting a $\beta$ model (\eg Sarazin 1986)
to the azimuthally averaged synthetic X-ray surface brightness at the given projection.
A representative velocity dispersion and the fitted core radii can be seen in Table \ref{tab1.3}.

In order to more closely mimic the observational analysis, we have chosen X-ray morphologies
that appear nominally relaxed (see Figure \ref{depro}). In those images containing obvious substructure such
as multiple peaks, we subtracted the secondary peaks from the image before obtaining the X-ray surface brightness profile.  This
is similar to the analysis done by Henry \etal (1993) on A2256 in which the region of the X-ray image containing the subcluster
peak was simply avoided.

The results of the deprojection can be seen in Figures \ref{depdens} and \ref{deptemp}. In Figure \ref{depdens}, we display
the true and deprojected gas density profiles for each epoch. As one can see, the gas density is
well constrained by the X-ray surface brightness. Errors are typically small and localized
in the core where resolution effects (\ie smoothing of the profiles) are most significant.
In Figure \ref{deptemp}, the discrepancies between the
two temperature profiles can often be large particularly in the core where they can approach a factor
of two or more. These errors result from both the non-equilibrium conditions and the systematic
errors of the deprojection discussed in \S \ref{initmass}. 

Figure \ref{depmass} shows the mass profiles derived from the deprojected
temperature and density profiles overlaid with the true mass profiles.
As expected, the errors are considerably larger than those obtained from
the true profiles (\S \ref{truemass}).  Peak-to-peak errors approach a factor of 3.

In the early merger examples shown here, the mass estimates
are systematically high by 50\% and $>$100\% at $-0.25$ and 0.0 Gyr, respectively. 
These errors are largely due to an overestimation of the velocity dispersion which is contaminated by the 
infall velocity of the subcluster. In the examples
presented here, the effect is enhanced because the merger axis is projected out of the plane
of the sky. In Figure \ref{depmass}b, the projection is greater than in Figure \ref{depmass}a, and the error in 
the mass determination is correspondingly larger.

In the examples depicting the period shortly after core passage (Figure \ref{depmass}c and d), there do
not appear to be any strong systematic errors in the mass estimates. This may result
because two competing effects are cancelling out. In both Figures \ref{depmass}c and \ref{depmass}d, the
X-ray core radii have increased relative to the dark matter core. This should result in
a underestimation of the central temperature and mass. However,
this effect is compensated for by the overestimation of the velocity dispersion which
raises the temperature and mass estimate. Projection effects can be severe during this
period. It is difficult to generalize, but merger remnants in the plane of the sky will tend 
to have their masses underestimated by the standard deprojection method
while those significantly out of the plane will tend to have their masses overestimated.

For $t>$2 Gyrs, the examples presented here tend to have their masses systematically underestimated. 
In both Figures \ref{depmass}e and \ref{depmass}f, the X-ray
core radii are significantly larger than the dark matter core radii (nearly a factor of 2 in \ref{depmass}f) causing the central 
temperatures and masses to be severely underestimated. Because of the prolate structure of the merger
remnant, the X-ray cores are subject to projection effects such that they will appear larger when viewed 
perpendicular to the merger axis. Projection effects in the velocity dispersion
 tend to be minimized during the later merger stages.
There is some residual anisotropy in the velocity dispersion ($\sim$5-15\%) even after 5 Gyrs, but it is small 
compared to the velocity anisotropy early in the merger ($\sim$25-50\%). 

Figure \ref{depmerr1} shows the evolution of the mean emissivity-weighted error in the deprojected mass estimate.
We have averaged the fractional discrepancy between the deprojected and true mass profiles for $r>$0.25 Mpc and weighted 
it by the X-ray surface brightness at that radius thereby accentuating uncertainties in the brighter
central regions of the clusters where one is more likely to make observational mass estimates.
Often, we find post-merger mass estimates systematically low by $\sim$50\%.
As a comparison, we have also included the mean absolute error at the same epochs. The mean is taken over the same range in radius, but is not emission-weighted. Comparison with Figure \ref{eqmerr1} shows that the mass errors are not
so exclusively linked to the gas dynamics as in \S \ref{truemass}  This would seem to imply that significant systematic effects are
being introduced by the deprojection technique.
The mean emissivity-weighted errors show a trend toward significant underestimation of the mass even after
5 Gyrs. The discrepancy between the mean absolute errors and the mean emissivity-weighted errors shows
that errors are typically larger at small radii. As discussed above, the inflated X-ray cores cause
a systematic underestimation of the central temperature and an underestimation of the mass.

\subsection { Isothermal Spectroscopic Temperature}
\label{isomass}

Since the errors in the mass estimates are tied so closely to the gas temperatures, it
makes sense to use spectroscopically determined temperatures rather than the
deprojection based temperatures whenever possible. However, even the spectroscopically
determined temperatures are poorly determined (\eg David \etal 1993), and the limited spatial resolution requires
the assumption of isothermality. Since merging clusters are not isothermal (\S \ref{tmorph}) we need to assess the
uncertainties introduced by that assumption. We have therefore recalculated
the mass estimates as before, using the deprojected density profiles, although this time we  have replaced
the deprojected temperature profiles with mean emissivity-weighted temperatures. This method  helps to reduce some of the systematic errors
noted previously. However,  since the density and 
temperature profiles are no longer self-consistent, considerable noise is added to the derived mass profiles.

In Figure \ref{depmerr2}, we have plotted the same quantities as in Figure \ref{depmerr1} only now for the isothermal analysis.
We find that the mean emissivity-weighted errors have been reduced significantly by inclusion
of the improved temperature estimates. The errors are always less than 50\% and for $>$2 Gyrs
after core passage they
are less than 25\%. We also note that the mean absolute errors are, at least for $r >$0.25 Mpc, somewhat
larger than in Figure \ref{depmerr1}. This is due to the increased noised induced by restricting the mass
calculation to isothermal conditions.  Peak-to-peak errors, can be quite large. Typically, they are of
order 50-100\%, but in extreme cases, they can exceed a factor of 3.

\subsection{ A Comparison with Previous Work}

Schindler (1996) quotes an RMS uncertainty of 15\% for mass estimates
within the central 2 Mpc of a cluster  for her merger simulations. However, she does note several instances where the errors can be
considerably larger. In the case of isothermal clusters, deviations are between
20\% and 35\%. Near the radius of shocks, cluster masses can be overestimated
by more than 100\%, and substructure can cause severe underestimations
at the radius of the substructure. She also notes that in some
instances errors in the mass estimates can be greater at radii $<$0.5 Mpc
(20-30\%) than at larger radii (10-20\%). 

We find that these results are in general agreement with our own although our uncertainties tend to be somewhat larger,
and we find an epochal dependence in the errors. Our quoted errors tend to be  larger because we used deprojected density {\it and} temperature profiles (\S 
\ref{depromass}) or mean isothermal temperatures (\S \ref{isomass}) whereas she used the deprojected density profile along with the true temperature profile. Also, we have weighted our 
errors by the X-ray 
surface brightness at the given radius thus emphasizing the more observationally
important inner regions of the cluster. Most clusters do not have
X-ray emission detected beyond 0.5-1.0 Mpc let alone 2-3 Mpc. Finally,
we find the greatest uncertainty in mass estimates made within the
first 2 Gyrs after core passage when equilibrium conditions are most
severely violated.

\subsection { Implications for the Baryon Fraction}
\label{bfract}

Uncertainty in the baryon mass fraction has two sources. They are the uncertainty in the total mass estimate,
as described above, and the uncertainty in the total gas mass. The total gas mass can be obtained simply
by integrating the deprojected gas density profiles. The similarity between the deprojected
density profiles and the true profiles (see Figure \ref{depdens}), would seem to indicate that the total
gas mass is well constrained by the X-ray emission.  We find errors in the
integrated gas mass to be typically of order 5\%. There are a few exceptions when errors exceed 10\%
at a given radius. 

Since the uncertainties in the total gas masses are small relative to those in the total mass, the primary
source of uncertainty in the baryon fraction resides in the total mass estimate. Figure \ref{bafract}
shows the ratio of the observed (\ie deprojection-based) baryon fraction and the true baryon
fraction (solid line) {\it within} a given radius for the deprojection sample. 
As a comparison of the relative
uncertainties, we have also included the ratio of the integrated deprojection-based gas mass to the 
true gas mass (dotted line). In both cases, a ratio of unity indicates an agreement
between ``observations" and ``truth." 

In the examples of the early merger
stages (Figure \ref{bafract}a and b), the baryon fraction is systematically underestimated by $\sim$50\%. Recall that the
total mass was overestimated. In the examples of the later merger states (Figure \ref{bafract}c, e, and f), 
there is a systematic overestimation of the baryon fraction, particularly
in the central regions ($r<$1.0 Mpc). The degree of overestimation, often greater than 50\%, could help to reduce the discrepancy between observed and theoretically derived baryon fractions.
As an example, White \& Fabian (1995) find the baryon fractions in a sample of
13 clusters to range between 0.1-0.22 within 1 Mpc while Big Bang nucleosynthesis calculations predict a value of $\sim$0.06
for an $\Omega$=1 universe (Olive \etal 1990).  Both the observed and theoretical values assume H$_\circ$=50 \kms Mpc$^{-1}$.
It should be kept in mind that much of the discrepancy observed in Figure \ref{bafract} is due to the systematics
of the deprojection technique as revealed in the mass estimates. If we calculate the baryon fraction
as a function of radius using the
deprojected gas density and the total mass estimates derived as in \S \ref{truemass} (\ie using true density and temperature profiles), then the RMS uncertainties in the
baryon fraction are typically less than 15\% except within the first 1 Gyr after core passage
in which case they can exceed 40\%.

\subsection { Virial Mass Estimates}
\label{viral}

As an illustrative example of the relative errors involved in optical
velocity-based mass estimates, we have calculated the virial masses of each cluster system in the deprojection sample.  The virial mass is given by
\begin{equation}
M_{VT}=\frac{3\pi}{G} \sigma_{R}^{2} \left\langle \frac {1}{r}\right\rangle^{-1}
\end{equation}
where $\sigma_{v}$ is the LOS velocity dispersion, $<$1/{\it r}$>$$^{-1}$ is the harmonic mean radius (\eg Oegerle \& Hill 1994),
and $G$ is the gravitational constant. It is important to remember
that $M_{VT}$ is the total cluster mass whereas the previous X-ray
analysis provided the mass enclosed within a given radius (eq. \ref{eqmass4}). In this analysis,
we calculated the velocity dispersion and harmonic mean radius for 100 draws of 100 randomly selected particles within 2.5 Mpc of the projected center of mass.
Applying this analysis to the initial, virialized primary cluster 
(4:1 mass ratio), we found $M_{VT}$=3.9$\pm$.5 $\times$ 10$^{14}$ 
M$_\odot$ compared to a true mass of 4.0 $\times$ 10$^{14}$ M$_\odot$ for
a 2.5\% error.  The results of the same analysis applied to the
deprojection sample can be seen in Table \ref{tab2.3}.

From Table \ref{tab2.3}, it is apparent 
that the errors are greatest when the velocity dispersion is most strongly
contaminated by the infall velocity. This occurs early in the mergers (near core passage),
when the merger axis is projected out of the plane of the sky, and for
lower mass ratio mergers where the probability of drawing a subcluster
particle is higher.  The minimum errors occur for high mass ratio mergers,
mergers in the plane of the sky, and in the later stages of the merger
when the system is returning to equilibrium. This result differs from
that of Bird (1995) who finds that correcting for substructure in virial
mass estimates more strongly affects the mean galaxy separation rather
than the velocity dispersion.

Observationally, there are
more sources of error than are addressed in this analysis. We do not
include contamination by foreground and background galaxies, a 
galaxy/dark matter velocity bias, nor do we include
measurement errors in the individual particle velocities. It should also be noted
that we do not attempt to remove obvious substructure which could
somewhat reduce the errors. Identification of substructure is strongly dependent
on a complete sampling of both the galaxy spatial and velocity distributions and is therefore
very difficult in most clusters.

\section {SUMMARY}
\label{summary}

In this paper, we have outlined the many observable consequences of cluster-subcluster 
mergers finding evidence of merger activity in a variety of cluster observations.
We have also investigated the effects of recent merger activity and the resulting
non-equilibrium conditions on the accuracy of equilibrium-based cluster mass estimates
and estimates of the cluster baryon fraction.

In the X-ray emission, we not only find evidence of obvious substructure (\ie multiple
peaks), but we also find more subtle evidence of non-equilibrium conditions such as 
isophotal twisting, centroid shifting
and the enlargement of the X-ray core relative to the dark matter distribution.
Elongation of the X-ray morphology is also a consequence of mergers although it is
relatively long-lived and therefore may not be a sensitive indicator of recent merger activity.

The X-ray emission also contains valuable temperature information. We find that one
of the strongest indicators of recent merger activity is in the temperature morphology of the X-ray emitting gas. Mergers tend to result in rather large temperature
inhomogeneities ($>$few keV) on scales $<$0.5 Mpc. This result is quite consistent with
recent high resolution temperature maps of several rich clusters (\eg Briel \& Henry 1994;
Henry \& Briel 1995, Markevitch \etal 1994; Arnaud \etal 1994), all of which have other signatures of
recent merger activity.

The gas dynamics also result in several observable consequences. We have presented a LOS
velocity map as an example of what might become available with the launch of {\it ASTRO-E}. 
The merger results in a very complex velocity structure that persists for several Gyrs
after the merger. The specifics of the observational analysis need to be addressed in
more detail in future work.  

We have added quantitative support
to the idea that bulk gas flows resulting from mergers can produce the wide-angle tailed (WAT)
radio source morphology. Although not specifically addressed in this work, we also believe
that bulk flows may result in the disruption of cooling flows and that this is true
even in high mass ratio mergers. McGlynn \& Fabian (1984) proposed that mergers of
nearly equal mass clusters might disrupt cooling flows. We also expect mergers to
disrupt the H$\alpha$ filaments which are often associated with cooling flows (see Baum 1992) and thus
may explain the results of Donahue, Stocke, \& Gioia (1992) which show a reduction in the number
of cooling flows (as indicated by H$\alpha$ filaments) in recent epochs. We also link mergers,
via shocks, turbulence and bulk flows, to the formation of Mpc-scale radio halos. We further contend,
based on the decay time for the ICM dynamics, that halos may be relatively long-lived compared
to the canonical synchrotron lifetime of 10$^8$ years.

Mergers may also influence the evolution of individual galaxies. As described here,
mergers generate expansive shocks in the ICM through which individual galaxies may pass. In doing so,
they will experience an increase in ram pressure which can induce a burst of star formation
changing the observed galaxy  colors and stellar content. The ram pressure may also
strip the galaxy of its gas content reducing the potential for future star formation and thus
altering the galaxy's observed color and morphology.

The dynamics induced by the merger were shown to affect the equilibrium-based
cluster mass estimates. Even in a best case scenario, (\ie knowledge of the true
temperature and density profiles), RMS errors in the mass estimates can approach 40\%
in the first 2 Gyrs after core passage because of shocks and bulk flows in the ICM. At later times, RMS errors are reduced to
$\sim$10\%. Local errors in the mass estimates, particularly in the early stages
can be in excess of 100\%.  Using density and temperature profiles obtained from the standard
X-ray deprojection technique adds potentially severe systematic errors. These can be
somewhat reduced by using a spectroscopically derived temperature rather than the deprojected
temperature profile. Still, the discrepancies in the mass estimates can be large, approaching
50\% within 2 Gyrs of core passage and 25\% for later epochs. These results are consistent
with those of  Evrard (1994) and somewhat larger than those of Schindler (1996). 

Of course, it should be recognized that
we have only analyzed a limited sample of merger remnants and extrapolation to the Universe
at-large is rather tricky. First, the uncertainties can vary markedly for any given cluster depending on the
details of the analysis and the radial extent of the X-ray data since there is a radial dependence on the fractional mass
discrepancies. Also, projection effects can be severe. Second, the epochal dependence on the mass estimate
discrepancies requires knowledge of merger rates. That is, if
major mergers occur more often than every 2 Gyrs, then finding a truly relaxed cluster
may be very difficult. It has been suggested by Edge, Stewart \& Fabian (1992) that any given cluster
will experience a merger every 2-4 Gyrs.

The errors in the mass estimates translate directly into errors in the observed
baryon fraction within clusters of galaxies. The tendency to underestimate the total mass
throughout the post-merger phase may account for some of the discrepancy between the expected and observed 
baryon fractions in some clusters.

\bigskip
This work was funded by a NASA Long Term Space Astrophysics Grant NAGW-3152 
and NSF grant AST-9317596 to JOB and CL. KR would like to thank the New Mexico NASA
Space Grant Consortium for partial support. We thank A. Klypin, J. Stocke, J. Pinkney,
M. Ledlow, P. G\'{o}mez, and Ren\'{e} Walterbos for useful discussions and
comments.

\newpage

\section*{REFERENCES}
 \everypar=
   {\hangafter=1 \hangindent=.5in}

Arnaud, K. A. 1988, in Cooling Flows and Clusters of Galaxies, eds. Fabian, A.C., (Kluwer Academic Publishers) 31.

Arnaud, K. A., Mushotsky, R. F., Ezawa, H., Fukazawa, Y., Ohashi, T., Bautz, M. W., Crewe, G. B., Gendreau, K. C., Yamashita, K., Kamata, Y., \& Akimoto, F. 1994, ApJ, 436, L67

Baum, S. A. in Clusters and Superclusters of Galaxies, eds. Fabian, A. C. NATO ASI series (The Netherlands: Kluwer), 171

Bird, C. M. 1995, ApJ, 445, 81L

Briel, U. G., Henry, J. P., Schwarz, R. A., B\"ohringer, Ebeling, H., Edge, A. C., Hartner, G. D., Schindler, S., Tr\"umpler, J., \& Voges, W. 1991, A\&A, 246, L10.

Briel, U. G., \& Henry, J. P. 1994 Nature, 372, 439

Bryan, G. L., Cen, R. Y., Norman, M. L., Ostriker, J. P., \& Stone, J. M. 1994, ApJ, 428, 405

Burns, J. O., Gregory, S. A., O'Dea, C. P., \& Balonek, T. J. 1986, ApJ, 307, 73

Burns, J. O., Roettiger, K., Ledlow, M., \& Klypin, A. 1994, ApJ, 427, L87

Burns, J. O., Roettiger, K., Pinkney, J., Perley, R. A., Owen, F. N., \& Voges, W. 1995, ApJ, 446, 583

Butcher, H., \& Oemler, A. J. ApJ, 1978, 219, 18

Caldwell, N., Rose, J. A., Sharples, R. M., Ellis, R. S., \& Bower, R. G. 1993, AJ, 106, 473

Clarke, D. A., 1990, BAAS, 22, 1302

David, L., Slyz, A., Jones, C., Forman, W., Vrtilek, S. D., \& Arnaud, K. A. 1993, ApJ, 412, 479

Donahue, M., Stocke, J. T., \& Gioia, I. M. 1992, ApJ, 385, 49

Dressler, A., \& Gunn, J. E. 1983 ApJ, 270, 7

Edge, A. C., Stewart, G. C., \& Fabian, A. C. 1992, MNRAS, 258, 177

Edge, A. C., Stewart, G. C., Fabian, A. C., \& Arnaud, K. A. 1990, MNRAS, 245, 559

Eilek, J. A., Burns, J. O., O'Dea, C. P., \& Owen, F. N. 1984, ApJ, 278, 37

Eilek, J. A., \& Hughes, P. E. 1990, in Astrophysical Jets, ed. P. E. Hughes (Cambridge Univ. Press), 428

Evrard, A. E. 1990, ApJ, 363, 349

Evrard, A. E. 1991, MNRAS, 248, 8

Evrard, A. E. 1994, in  Clusters of Galaxies, Proceedings of the XIVth
Moriond Astrophysics Meeting, Editions Frontieres, Gifsur-Yvette, eds. Durret, F., Mazure, A. \& Tr\^{a}n Thanh V\^{a}n, 241

Fabricant, D., Lecar, M., \& Gorenstein, P. 1980, ApJ, 241, 552

Feretti, L., Perola, G. C., \& Fanti R. 1992, A\&A, 265, 9

Frenk, C. S., Evrard, A. E., White, S. D. M., \& Summers, F. J 1995, preprint

Giovannini, G., Feretti, L., Venturi, T., Kim, K. -T., \& Kronberg, P. P. 1993, ApJ, 406, 399.

Girardi, M. Biviano, A., Giuricin, G., Mardirossian, F., \& Mezzetti, M. 1993, ApJ, 404, 38

G\'{o}mez, P. Pinkney, J., Burns, J. O., Wang, Q., \& Owen, F. N. 1996, ApJ, submitted

Gunn, J. E., \& Gott, J. R. 1972, ApJ, 176, 1

Hanisch, R. J. 1982, A\&A, 116, 137

Henricksen, M. J. 1993, ApJ, 414, L5

Henry, J. P., \& Arnaud,  K. A. 1991, ApJ, 372, 410

Henry, J. P., \& Briel, U. G. 1995, ApJ, 443, L9

Henry, J. P., Briel, U. G., \& Nulsen, P. E. J. 1993, A\&A, 271, 413

Herbig, T., \& Birkinshaw, M. 1994, BAAS, 26, 1403

Hernquist, L. 1987, ApJS,  64, 715

Jaffe, W. 1992, in Clusters and Superclusters of Galaxies, eds. Fabian, A. C. NATO ASI series, 109 

Kneib, J.P., Mellier, Y., Pell\'{o}, R., Miralda-Escud\'{e}, J., Le Borgne, J.-F., B\"{o}hringer, H., 
\& Picat, J.-P. 1995, A\&A, 303, 27

Loeb, A., \& Mao, S. 1994, ApJ, 435, 109L

Loken, C., Roettiger, K., Burns, J. O. \& Norman, M. L. 1995, ApJ, 445, 80

Markevitch, M., Yamashita, K., Furusawa, A., \& Tawara, Y. 1994, ApJ, 436, L71

McGlynn, T. A. \& Fabian, A. C. 1984, ApJ, 208, 709

Merritt, D., 1987, ApJ, 313, 121

Miley, G. K., Perola, G. C., van der Kruit, P. C., van der Laan, H. 1972, Nature, 237, 269

Miralda-Escud\'e, J., \& Babul, A. 1995, 449, 18

Mohr, J. J., Evrard, A. E., Fabricant, D. G., \& Geller, M. J. 1995, ApJ, 447, 8

Nakamura, F. E., Hattori, M., \& Mineshige, S. 1995, A\&A, 302, 649

Newberry, M., Boroson, T., \& Kirshner, R. 1990, ApJ, 350, 585

O'Donoghue, A., Eilek, J. A., \& Owen, F. N. 1993, ApJ, 408, 428

Oegerle, W. R., \& Hill, J. M. 1994, AJ, 107, 857

Oemler, A. 1992, in Clusters and Superclusters of Galaxies, ed. Fabian, A. C. NATO ASI Series (Kluwer Academic Publishers), 29

Olive, K. A., Schramm, D. N. Steigman, G., \ Walker, T. P. 1990, Phys. Lett. 236B, 454

Pinkney, J.,  1995, Ph.D. Thesis, New Mexico State University

Pinkney, J., Burns, J. O., \& Hill 1994, AJ, 108, 2031

Pinkney, J., Rhee, G. F., Burns, J. O., Hill, J. M., Oegerle, W.,
Batuski, D., \& Hintzen, P. 1993, ApJ, 416, 36

Raymond, J. C., \& Smith, B. W. 1977, ApJS,  35, 419

Roettiger, K., Burns, J. O., \& Loken, C. 1993, ApJ,  407, L53

Roettiger, K., Loken, C., \& Burns, J. O. 1996, ApJS, submitted

Roettiger, K. Burns, J. O., \& Pinkney, J. 1995, ApJ, 453, 634

R\"{o}ttgering, H, Snellen, I., Miley, G., de Jong, J. P., Hanisch, R. J., \& Perley, R. 1994, ApJ, 436, 654

Ruzmaikin, A. 1989, MNRAS, 241, 1

Sarazin, C. 1986, Rev. Mod. Phys.,  58, 1

Schindler, S. 1996, A\&A, 305, 756

Spitzer, L., \& Baade, W. 1951, ApJ, 113, 413

Stone, J. M., \& Norman, M. L. 1992, ApJS,  80, 791

Tribble, P. 1993, MNRAS, 263, 31

Tyson, J. A., Valdes, F., \& Wenk, R. A. 1990, ApJ, 349, L1

White, D. A., \& Fabian, A. C. 1995, MNRAS, 273, 72

Wu, X. 1994, ApJ, 436, L115

\newpage

\section*{Figure Captions}

{\bf Fig. \ref{evol}:} Synthetic X-ray surface brightness images of a 4:1 mass ratio
merger (the evolutionary sample, \S \ref{esamp}) at six epochs. The epoch is 
is listed above each panel and is relative to the time of core passage. The subcluster enters from the top of the
panel and proceeds downward. Contours are uniformly spaced in the logarithm and constant from
one panel to the next, with dynamic range $\log$(S$_{max}$/S$_{min}$)=5.4. Each panel is 6.5 Mpc on a side.

{\bf Fig. \ref{depro}:} Synthetic X-ray surface brightness of the deprojection sample (\S \ref{dsamp}). Panels a-f
correspond directly to Mergers 1-6 in Table \ref{tab1.3}. Each panel is 6.5 Mpc on a side. The contours are 
uniformly spaced in the logarithm and relative to the peak surface brightness in each panel. The dynamic range 
in surface brightness represented
in each panel is $\log$(S$_{max}$/S$_{min}$)=5.0, 5.6, 4.5, 4.6, 4.25, 4.55.

{\bf Fig. \ref{prof}:} The relative expansion of the gas core (dashed) with respect to the dark matter core (solid) as a result of heating
and gas motions generated during the 4:1 mass ratio merger. a) Initially, the gas and dark matter cores 
are of comparable size. b) At 5 Gyrs after core passage, the gas core has nearly doubled in size where the 
dark matter core is virtually unchanged.

{\bf Fig. \ref{ellipses}:} The result of fitting elliptical isophotes to the X-ray surface brightness 
image in Fig. \ref{depro}d. The dimensions of the box are 3 Mpc on a side. The merger axis runs vertically.
This figure shows the elongation of the X-ray distribution along the merger axis, a radially dependent
ellipticity, and a significant centroid shift of $\sim150$ kpc.

{\bf Fig. \ref{temp}:} The emissivity-weighted temperature distribution for the evolutionary sample (\S \ref{esamp}).
Solid contours are labeled with the temperature in keV. The X-ray surface brightness is shown as
dotted contours that are uniformly spaced in the logarithm over the same dynamic range as in
Figure \ref{evol}. Each panel is 4 Mpc on
a side.

{\bf Fig. \ref{temp}:} The emissivity-weighted temperature distribution for the deprojection sample (\S \ref{dsamp}).
Solid contours are labeled with the temperature in keV. The X-ray surface brightness is shown as
dotted contours that are uniformly spaced in the logarithm over the same dynamic range as in
Figure \ref{depro}. Panels a-f correspond directly to Mergers 1-6 in Table \ref{tab1.3}. Each panel is 4 Mpc on
a side.

{\bf Fig. \ref{texample}:} A slice in temperature (filled contours)
along the merger axis through the cores of a 2:1 mass ratio merger.
The merger axis runs vertically with the slightly cooler subcluster
at the top and moving downward. The epoch is 0.25 Gyrs before core
passage. The contours are evenly space between 0.5 and 8.5 keV. Black
designates low temperatures. White designates high temperatures. The
image dimensions are 6.5 $\times$ 11.2 Mpc.

{\bf Fig. \ref{velocity}:} The gas dynamics. a) The observable emission-weighted mean LOS velocity field for Merger 4 of the deprojection sample 
(\S \ref{dsamp}). b) The emission-weighted LOS velocity dispersion.  In both (a) and (b), the velocity contours (solid) are labeled with
the velocity in \kmsp In (a), the velocities are in the rest frame of the primary cluster's dark matter core, and negative velocities are approaching the observer. The dotted contours are the X-ray surface brightness. Both panels are 2 Mpc on a side.

{\bf Fig. \ref{vbulk1}:} The evolution of the peak gas velocity within
a 2 Mpc box centered on the dark matter density peak for mergers having
mass ratios of 1:1 (+), 2:1 ($\ast$), 4:1 ($\diamond$),
and 8:1 ($\triangle$).

{\bf Fig. \ref{vbulk2}:}  The evolution of the width of bulk flows through the cluster core for mergers having mass ratios of 1:1 (+), 2:1 
($\ast$), 4:1 ($\diamond$),
and 8:1 ($\triangle$). The flow width is defined as the
width at the primary dark matter core at which the flow velocity drops to 0.5 of the peak or at which the flow direction
varies from that of the merger axis by more than 25$^\circ$.

{\bf Fig. \ref{vbulk3}:} The evolution of the mean gas velocity within
a 2 Mpc box centered on the dark matter density peak for mergers having
mass ratios of 1:1 (+), 2:1 ($\ast$), 4:1 ($\diamond$),
and 8:1 ($\triangle$).

{\bf Fig. \ref{ram}:} The evolution of the median ram pressure experienced by the N-body particles during the 8:1 mass ratio merger. Subcluster particles ($\diamond$), on average, experience an impulsive increase in ram pressure near the time of core passage.
A smaller but still sizable fraction of primary cluster particles ($\triangle$) also experience a large increase in ram pressure as they pass through
shocks generated during the merger.

{\bf Fig. \ref{mach}:} Mach number contours at 0.5 Gyrs after core passage in
the 2:1 mass ratio merger. Shocks are identified as the two opposing C-shaped
structures (closely spaced contours). The cluster core is located between
these structures. The region depicted is 6.5 $\times$ 11.2 Mpc.

{\bf Fig. \ref{staticmass}:} Mass Estimates of the initial hydrostatic cluster. The solid line is the true mass profile. The
dotted line represents the mass profile derived using Eqn. \ref{eqmass4} and the true gas density and temperature profiles.
The dashed line represents the mass profile derived using Eqn. \ref{eqmass4} and the deprojection-based gas density
and temperature profiles.

{\bf Fig. \ref{syserr}:} A comparison of the systematic errors associated with the X-ray deprojection of an isothermal cluster in hydrostatic
equilibrium. a) Errors in the
temperature profile induced by errors in the core radius ($\circ$) and velocity dispersion ($\diamond$). 
Filled symbols represent 20\% overestimates of the quantity. Open symbols represent 20\% underestimates
of the quantity. The solid line represents the profile derived using the best estimates of the
core radius and velocity dispersion. The ``true" gas temperature is 4.6 keV. b) The corresponding errors induced in the gas density profile.

{\bf Fig. \ref{eqmass1}:} A comparison of the true (dashed) and equilibrium-base mass profiles (solid) for the
evolutionary sample (\S \ref{esamp}). The evolutionary sample is a 4:1 mass ratio merger at the six epochs
listed above each panel. All times are with respect to core passage. The equilibrium-based 
mass profile was derived using Eqn. \ref{eqmass4} and the true (azimuthally averaged) gas temperature and density profiles.

{\bf Fig. \ref{eqmerr1}:} The mean absolute ($\diamond$) and RMS errors ($\triangle$) in the equilibrium-based mass estimates
of the evolutionary sample (\S \ref{esamp}, Fig. \ref{eqmass1}) as a function of time relative to core passage. The fractional
error is averaged over radii from 0.25 to 2.0 Mpc.

{\bf Fig. \ref{depdens}:} A comparison of the true (solid) and deprojected (dotted) gas density profiles for the
deprojection sample (\S \ref{dsamp}). Panels a-f correspond directly to Mergers 1-6 in Table \ref{tab1.3}. All profiles 
have been normalized.

{\bf Fig. \ref{deptemp}:} A comparison of the true (solid) and deprojected (dotted) gas temperature profiles for the
deprojection sample (\S \ref{dsamp}). Panels a-f correspond directly to Mergers 1-6 in Table \ref{tab1.3}.

{\bf Fig. \ref{depmass}:} A comparison of the true (solid) and deprojection-based (dotted) mass profiles for the deprojection
sample (\S \ref{dsamp}). Panels a-f correspond directly to Mergers 1-6 in Table \ref{tab1.3}.

{\bf Fig. \ref{depmerr1}:} The mean emissivity-weighted error ($\diamond$) and mean absolute error ($\triangle$)
for the deprojection-based mass estimates (Fig. \ref{depmass}) as a function of time relative to core passage. 
Since various merger parameters are represented in this sample, this figure cannot be viewed as a true
time sequence although the magnitude of the uncertainties depicted here are considered to be representative of
the various epochs.  The fractional errors have been averaged between 0.25 and 2.0 Mpc for each profile in Fig. \ref{depmass}.

{\bf Fig. \ref{depmerr2}:} The mean emissivity-weighted error ($\diamond$) and mean absolute error ($\triangle$)
for the deprojection-based mass estimates in which the deprojected temperature profile has been 
replaced by the mean emissivity-weighted temperature. As in Fig. \ref{depmerr1}, this is not a true time sequence,
and the fraction error is averaged between 0.25 and 2.0 Mpc for each mass profile.

{\bf Fig. \ref{bafract}:} The error in the cluster baryon fraction based on the deprojected gas density profiles
and the equilibrium/deprojection-based mass profiles. The dotted line represents the uncertainty in the
baryon fraction assuming deprojection-based gas mass estimates and knowledge of the true mass profile. 
Unity implies a good agreement (dashed line). The solid line represents the uncertainty in the baryon 
fraction estimate assuming both total masses and gas masses based on deprojection. Panels a-f correspond
directly to Mergers 1-6 in Table \ref{tab1.3}.

\newpage

\onecolumn
\begin{table}

\begin{center}
\begin{tabular}{c c c c c c c c} 
\multicolumn{8}{c}{ Initial Cluster Parameters: Deprojection Sample}\\ \hline \hline
\multicolumn{1}{c}{ID} &
 \multicolumn{1}{c}{Mass Ratio} & 
 \multicolumn{1}{c}{Primary Mass} &
 \multicolumn{1}{c}{Epoch} &
 \multicolumn{1}{c}{P.A.} &
 \multicolumn{1}{c}{$\sigma_v$} &
 \multicolumn{1}{c}{$<$T$>$} &
 \multicolumn{1}{c}{$r_c$} \\
\multicolumn{1}{c}{ } &
 \multicolumn{1}{c}{ } & 
 \multicolumn{1}{c}{(10$^{14}$ M$_\odot$)} &
 \multicolumn{1}{c}{(Gyrs)} &
 \multicolumn{1}{c}{Degrees} &
 \multicolumn{1}{c}{ (\kms) } &
 \multicolumn{1}{c}{ (KeV) } &
 \multicolumn{1}{c}{ (kpc) } \\ \hline

1 & 2:1 & 2.0 & -0.25 & 45 & 790 & 2.8  & 245 \\
2 & 4:1 & 4.0 &  0.0  & 90 & 1300 & 8.1 & 162 \\
3 & 8:1 & 8.0 &  0.5  & 45 & 1055 & 9.5 & 411 \\
4 & 2:1 & 2.0 &  1.0  & 45 & 590 & 2.5  & 142 \\
5 & 4:1 & 4.0 &  2.0  &  0 & 660 & 4.5  & 333 \\
6 & 8:1 & 8.0 &  5.0  &  0 & 995 & 9.6  & 492 \\ \hline 
\end{tabular}
\end{center}
\caption[ Deprojection Sample Parameters]
{Column 1 is the merger ID. Column 2 is the cluster mass ratio. Column 3 is the primary cluster mass.
Column 4 is the merger epoch relative to the time of core passage. Column 5 is the projection angle of
the merger axis with respect to the plane of the sky. Column 6 is the LOS velocity dispersion. Column 7 is
the mean emissivity-weighted temperature within 1 Mpc of the X-ray maximum. Column 8 is core radius obtained
from fitting a $\beta$ model (eq. \ref{betamod})}
\label{tab1.3}
\end{table}

\begin{table} 
\begin{center}
\begin{tabular}{c c c } 
\multicolumn{3}{c}{ Virial Mass Comparison}\\ \hline \hline
\multicolumn{1}{c}{ID} &
 \multicolumn{1}{c}{True Mass} &
 \multicolumn{1}{c}{M$_{VT}$} \\
\multicolumn{1}{c}{ } &
\multicolumn{1}{c}{(10$^{14}$ M$_\odot$)} &
\multicolumn{1}{c}{(10$^{14}$ M$_\odot$)} \\ \hline

1 & 3.0 & 5.1$\pm$0.6   \\
2 & 5.0 & 10.0$\pm$0.15  \\
3 & 9.0 & 10.2$\pm$0.15 \\
4 & 3.0 & 4.0$\pm$0.7  \\
5 & 4.7 & 4.4$\pm$0.7   \\
6 & 8.6 & 8.3$\pm$1.1 \\ \hline 
\end{tabular}
\end{center}
\caption[Virial Mass Estimates ]
{Column 1 is the merger ID within the deprojection sample.  Column 2 is the true total mass. Column 3 is the
virial mass estimate.}
\label{tab2.3}
\end{table}
\newpage

\begin{figure}[htbp]
\centering \leavevmode
\epsfxsize=0.75\textwidth 
\vspace{7.5in}
\includegraphics{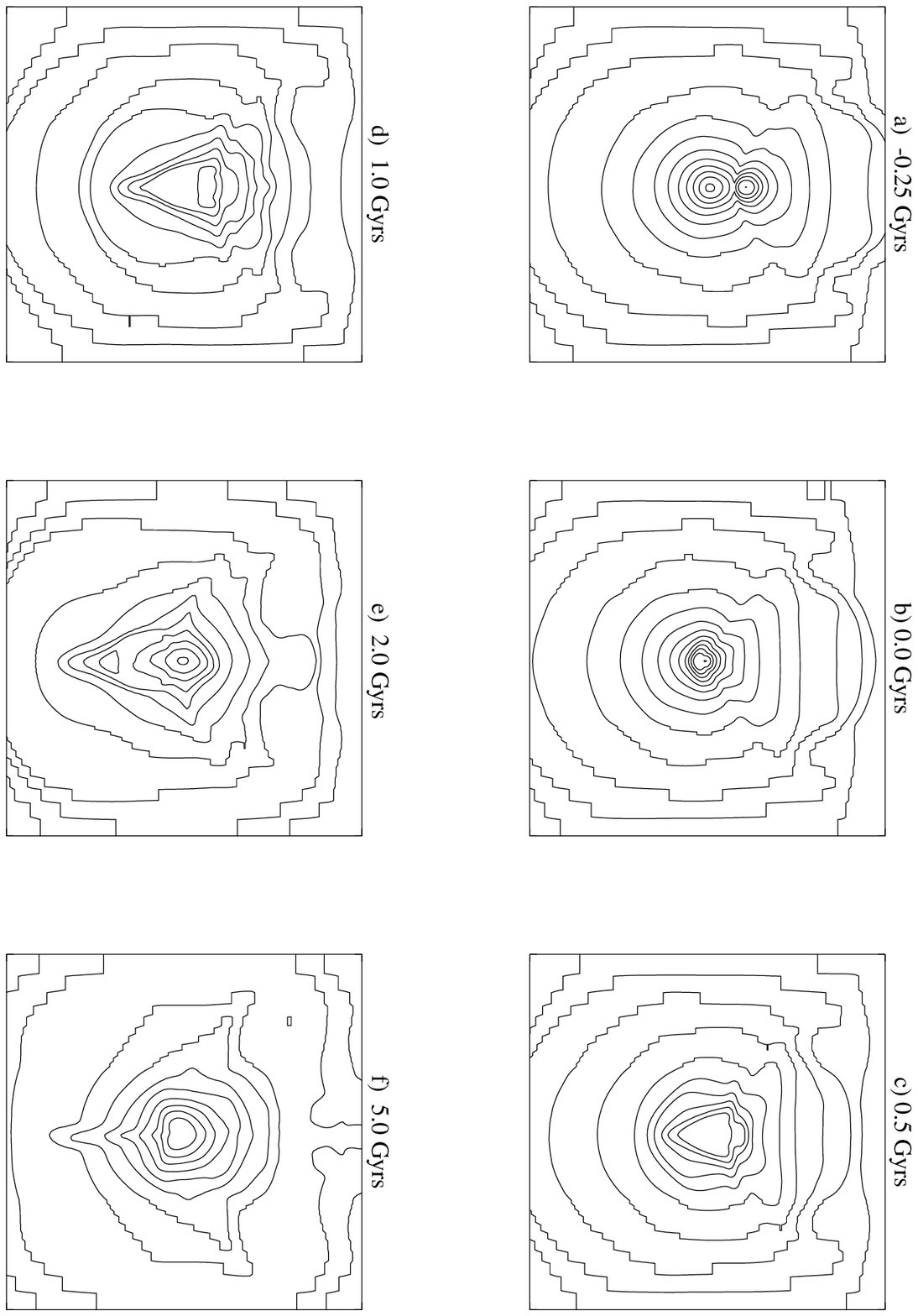}
\caption[Evolutionary Sample: Synthetic X-ray Images]
{   }
\label{evol}
\end{figure}

\begin{figure}[htbp]
\centering \leavevmode
\epsfxsize=0.9\textwidth 
\vspace{7.5in}
\includegraphics{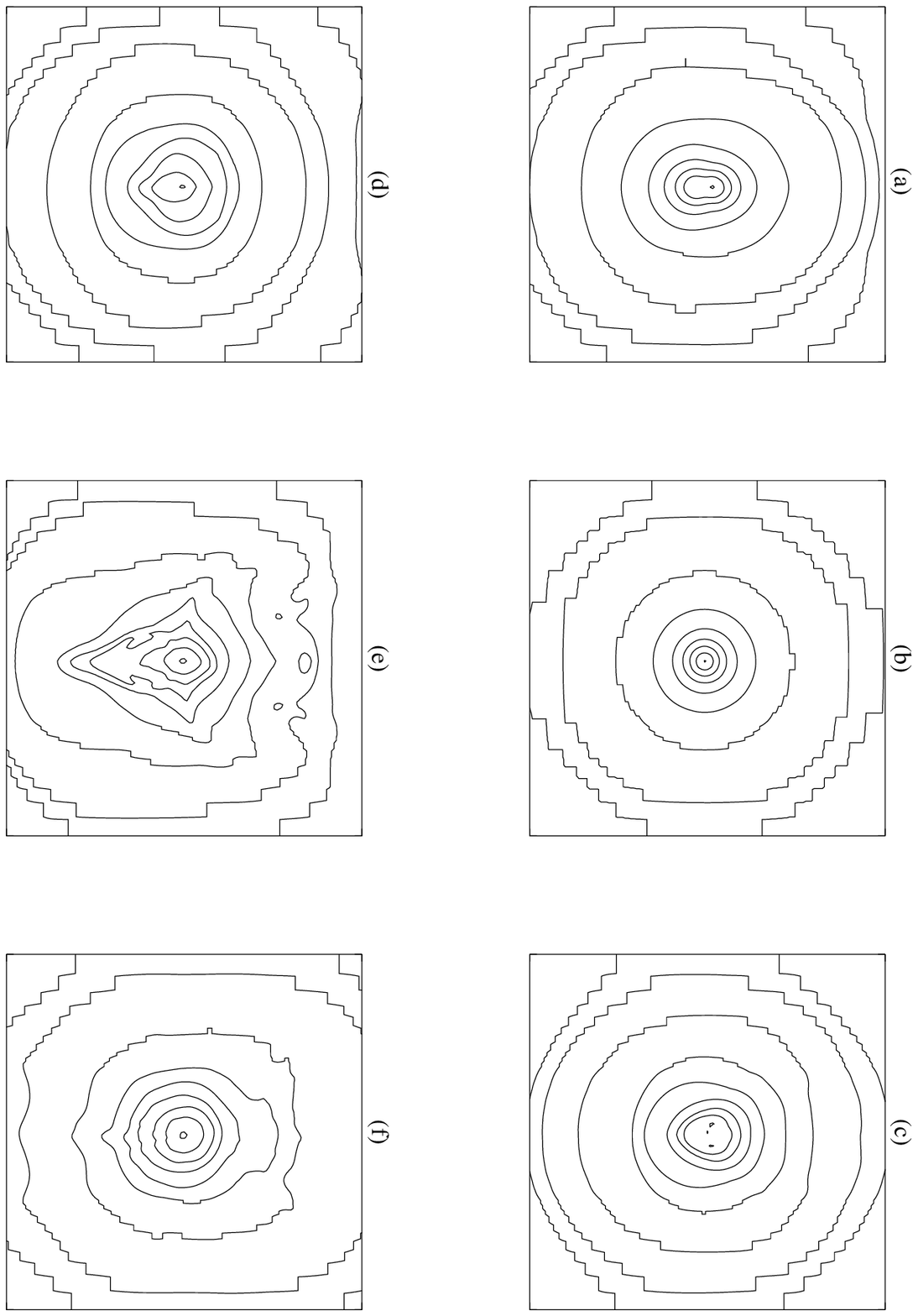}
\caption[Deprojection Sample: Synthetic X-ray Images]
{ }
\label{depro}
\end{figure}

\begin{figure}[htbp]
\centering \leavevmode
\epsfxsize=0.9\textwidth \epsfbox{ 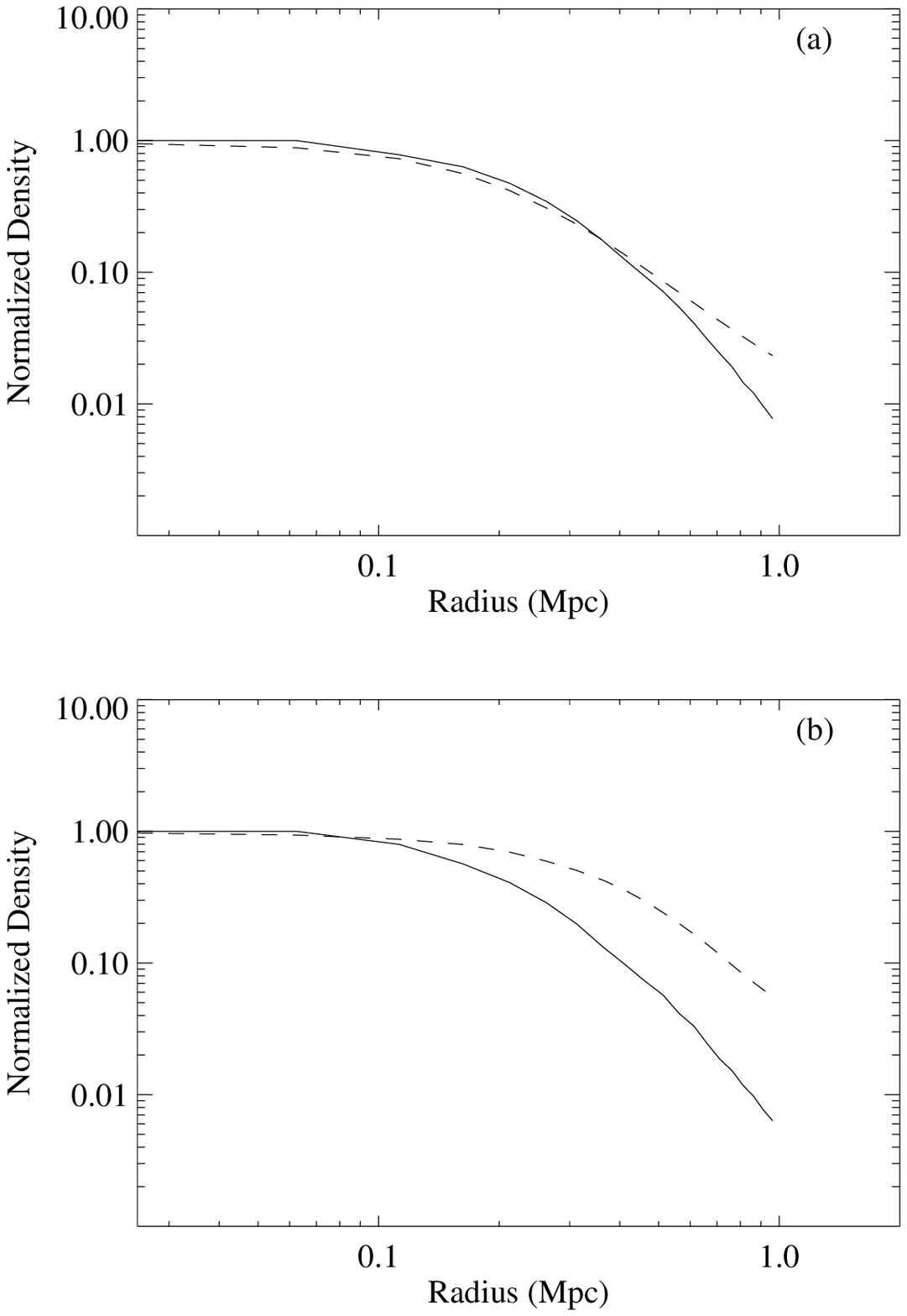}
\caption[Gas Core Expansion]
{ }
\label{prof}
\end{figure}

\begin{figure}[htbp]
\centering \leavevmode
\epsfxsize=0.7\textwidth \epsfbox{ 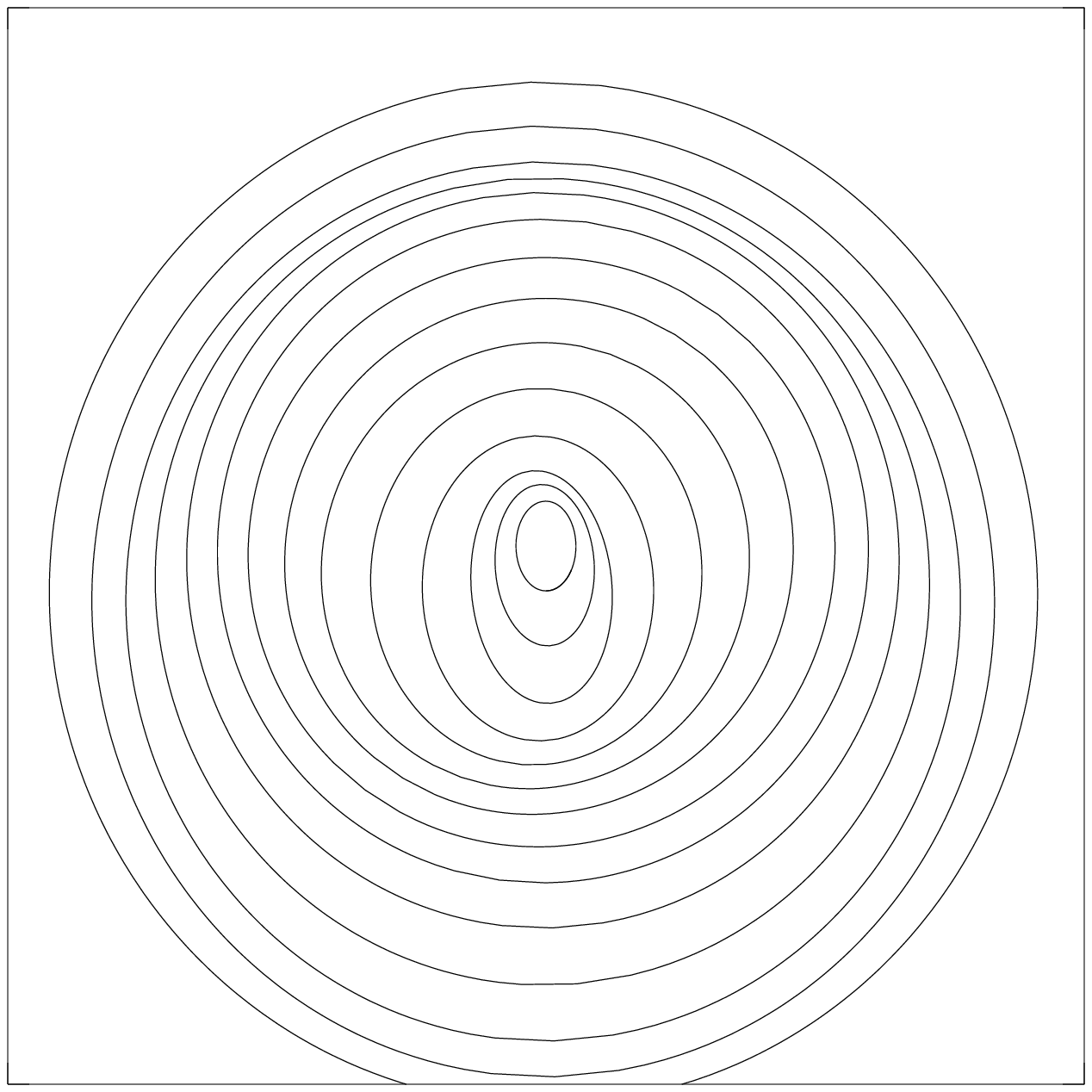}
\caption[Isophotal Fitting]
{  }
\label{ellipses}
\end{figure}

\begin{figure}[htbp]
\centering \leavevmode
\epsfxsize=0.9\textwidth 
\vspace{7.5in}
\includegraphics{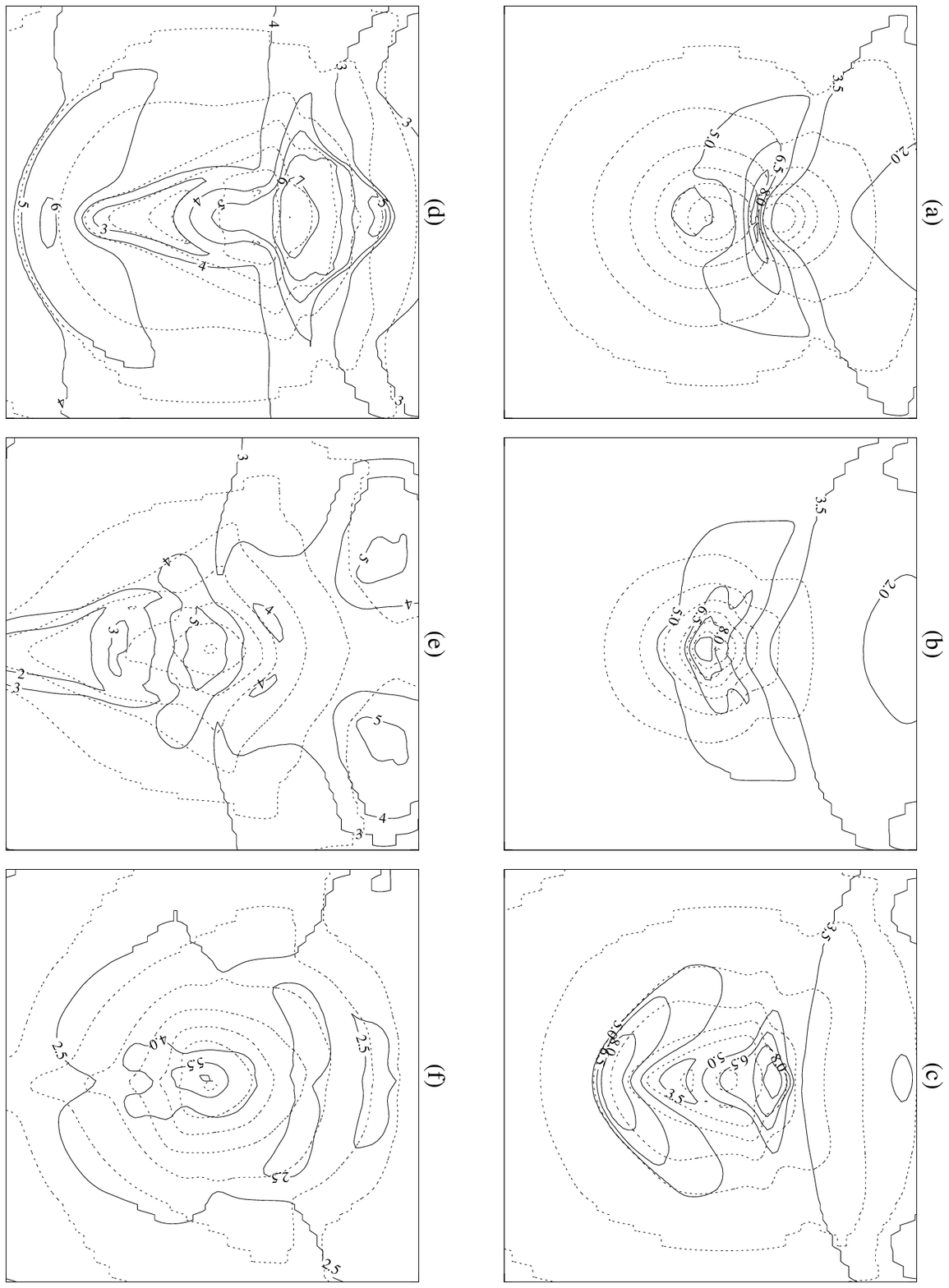}
\caption[Emission-Weighted Temperature Morphology]
{  }
\label{temp1}
\end{figure}

\begin{figure}[htbp]
\centering \leavevmode
\epsfxsize=0.9\textwidth 
\vspace{7.5in}
\includegraphics{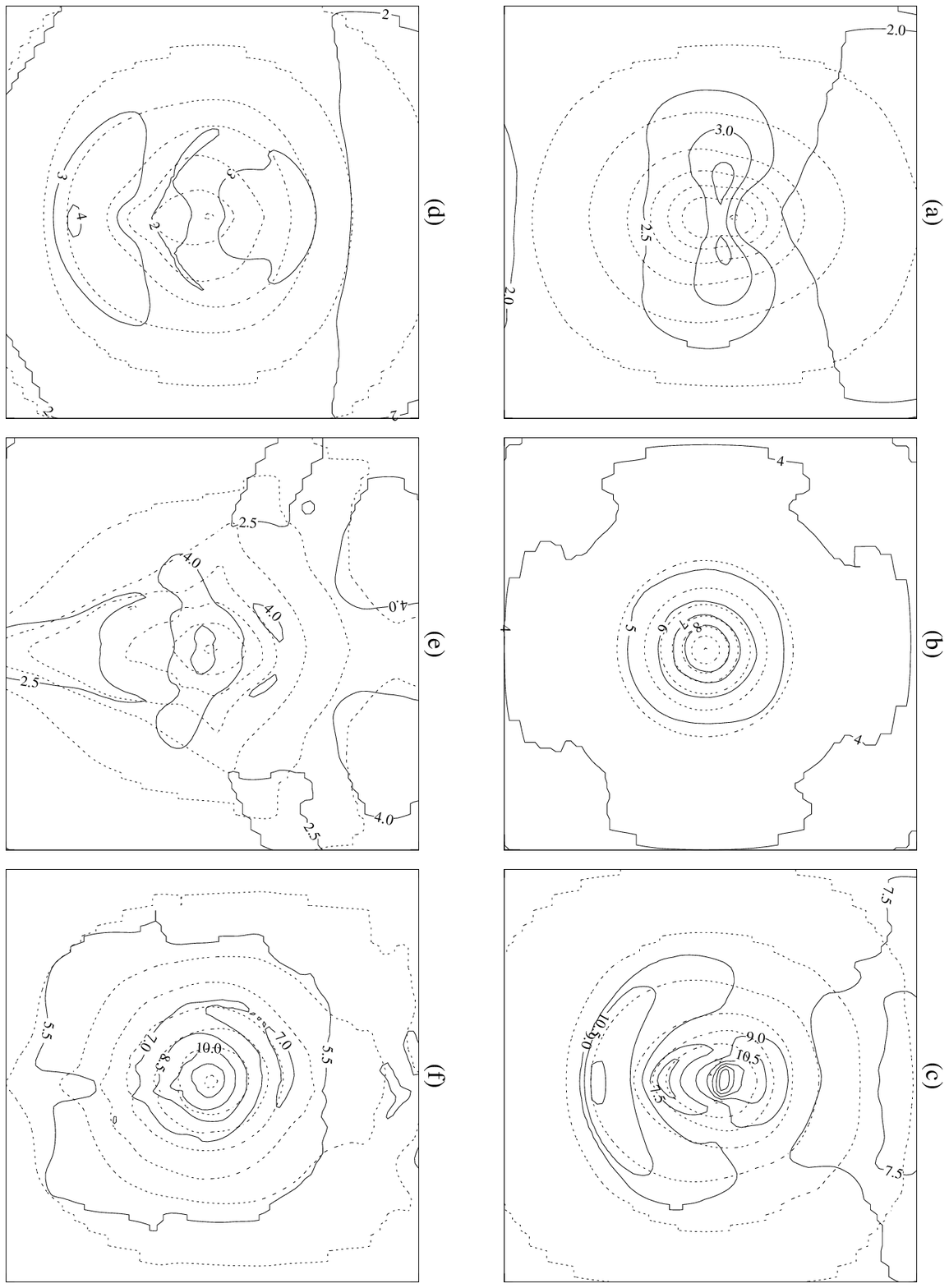}
\caption[Emission-Weighted Temperature Morphology]
{  }
\label{temp}
\end{figure}

\begin{figure}[htbp]
\centering \leavevmode
\epsfxsize=0.4\textwidth 
\epsfbox{ 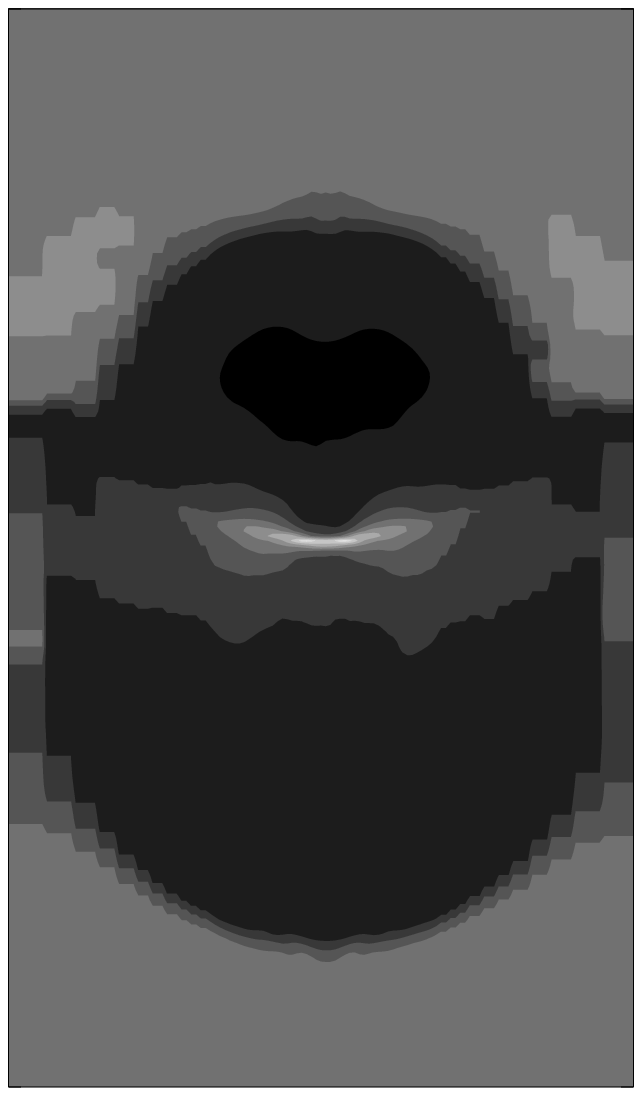}
\caption[true temp]
{  }
\label{texample}
\end{figure}

\clearpage
\begin{figure}[htbp]
\centering \leavevmode
\epsfxsize=0.5\textwidth \epsfbox{ 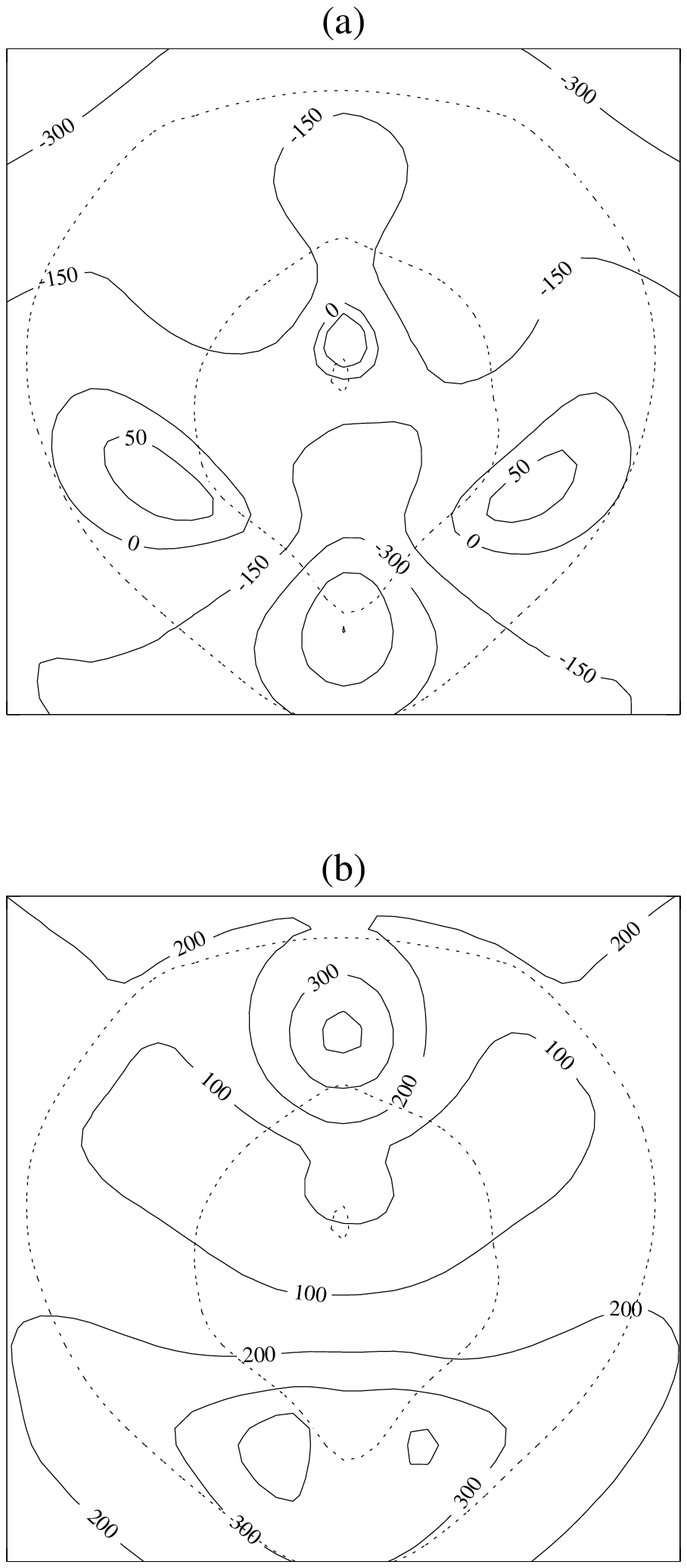}
\caption[Gas Velocity Field]
{ }
\label{velocity}
\end{figure}

\begin{figure}[htbp]
\centering \leavevmode
\epsfxsize=0.9\textwidth 
\epsfbox{ 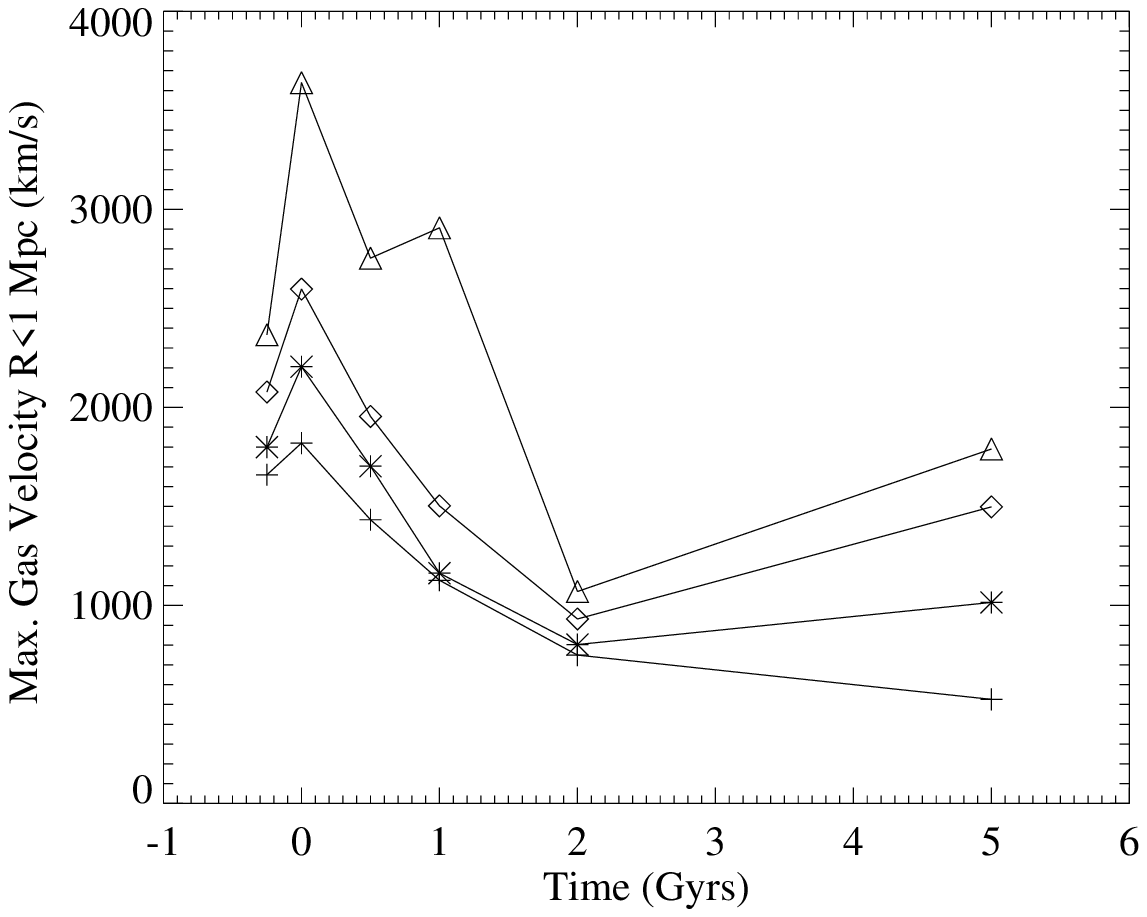}
\caption[bulk velocity peak]
{  }
\label{vbulk1}
\end{figure}

\begin{figure}[htbp]
\centering \leavevmode
\epsfxsize=0.9\textwidth 
\epsfbox{ 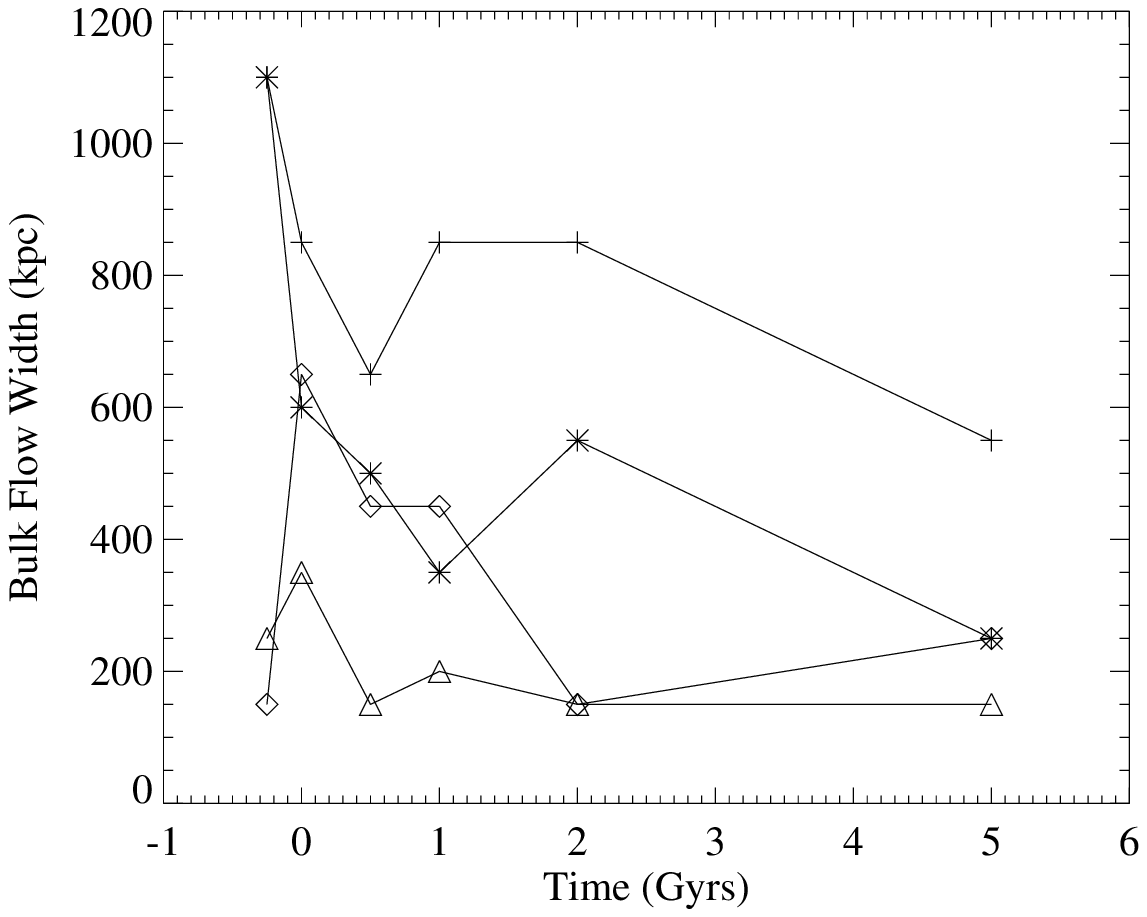}
\caption[bulk flow]
{  }
\label{vbulk2}
\end{figure}

\begin{figure}[htbp]
\centering \leavevmode
\epsfxsize=0.7\textwidth \epsfbox{ 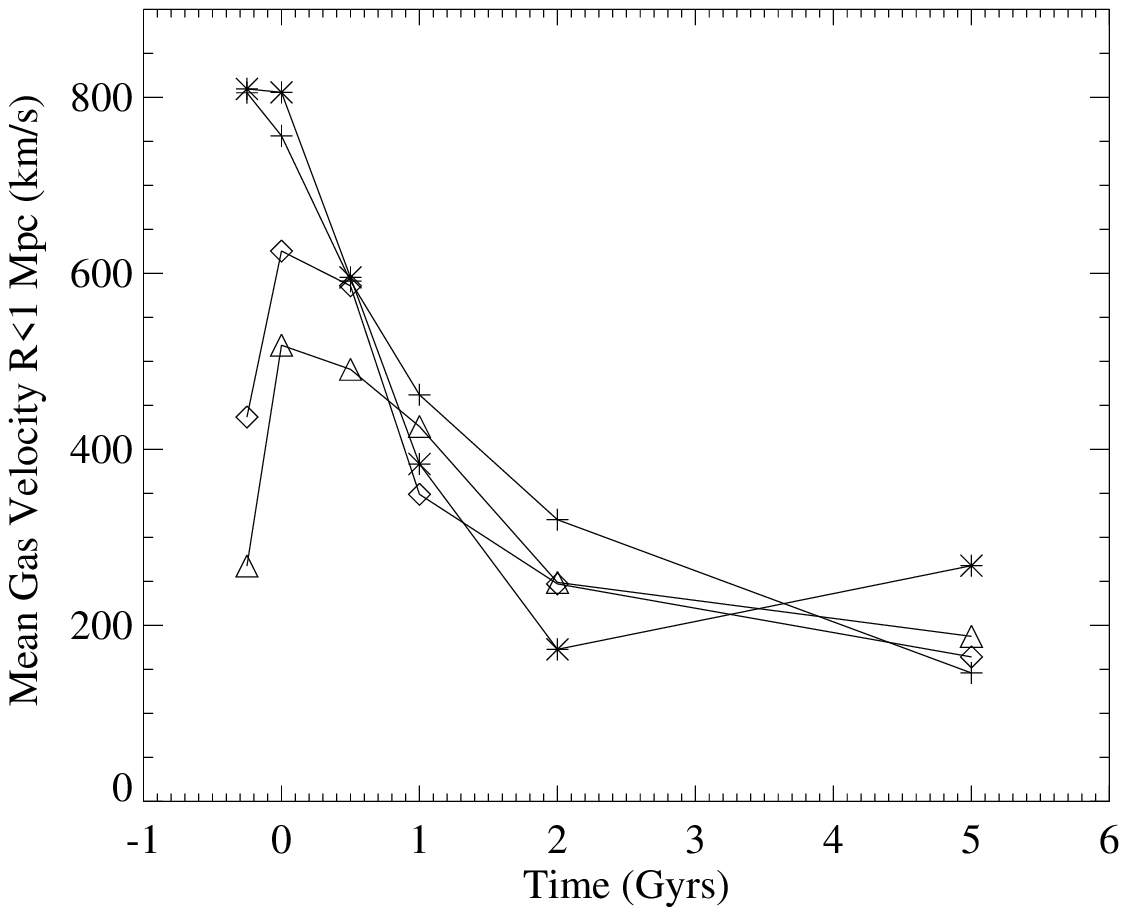}
\caption[ ]
{ }
\label{vbulk3}
\end{figure}

\begin{figure}[htbp]
\centering \leavevmode
\epsfxsize=0.6\textwidth \epsfbox{ 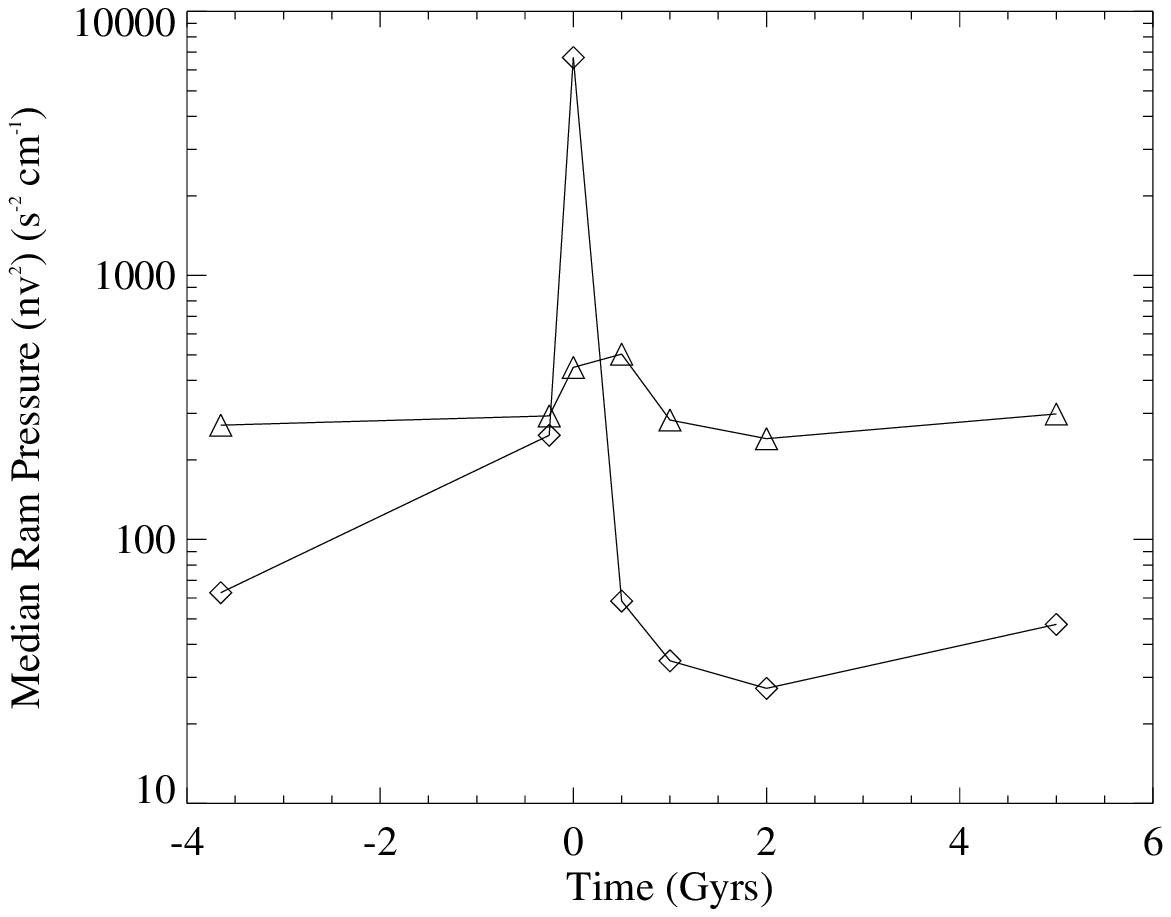}
\caption[Particle Ram-Pressure Evolution]
{ }
\label{ram}
\end{figure}

\begin{figure}[htbp]
\centering \leavevmode
\epsfxsize=0.4\textwidth \epsfbox{ 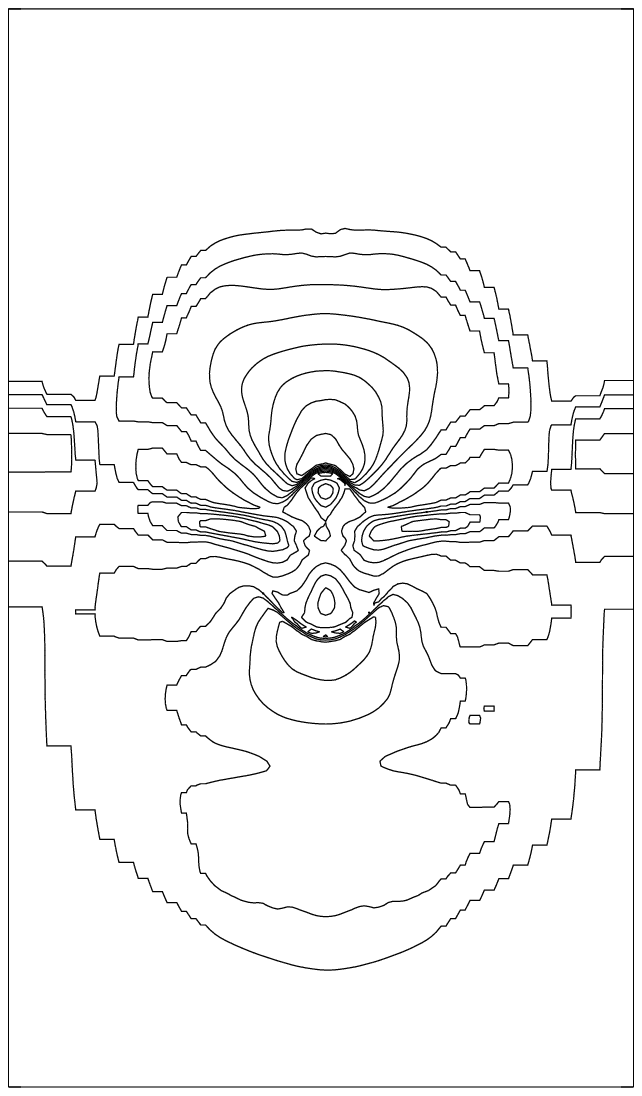}
\caption[Example of Mach number]
{ }
\label{mach}
\end{figure}

\begin{figure}[htbp]
\centering \leavevmode
\epsfxsize=0.6\textwidth \epsfbox{ 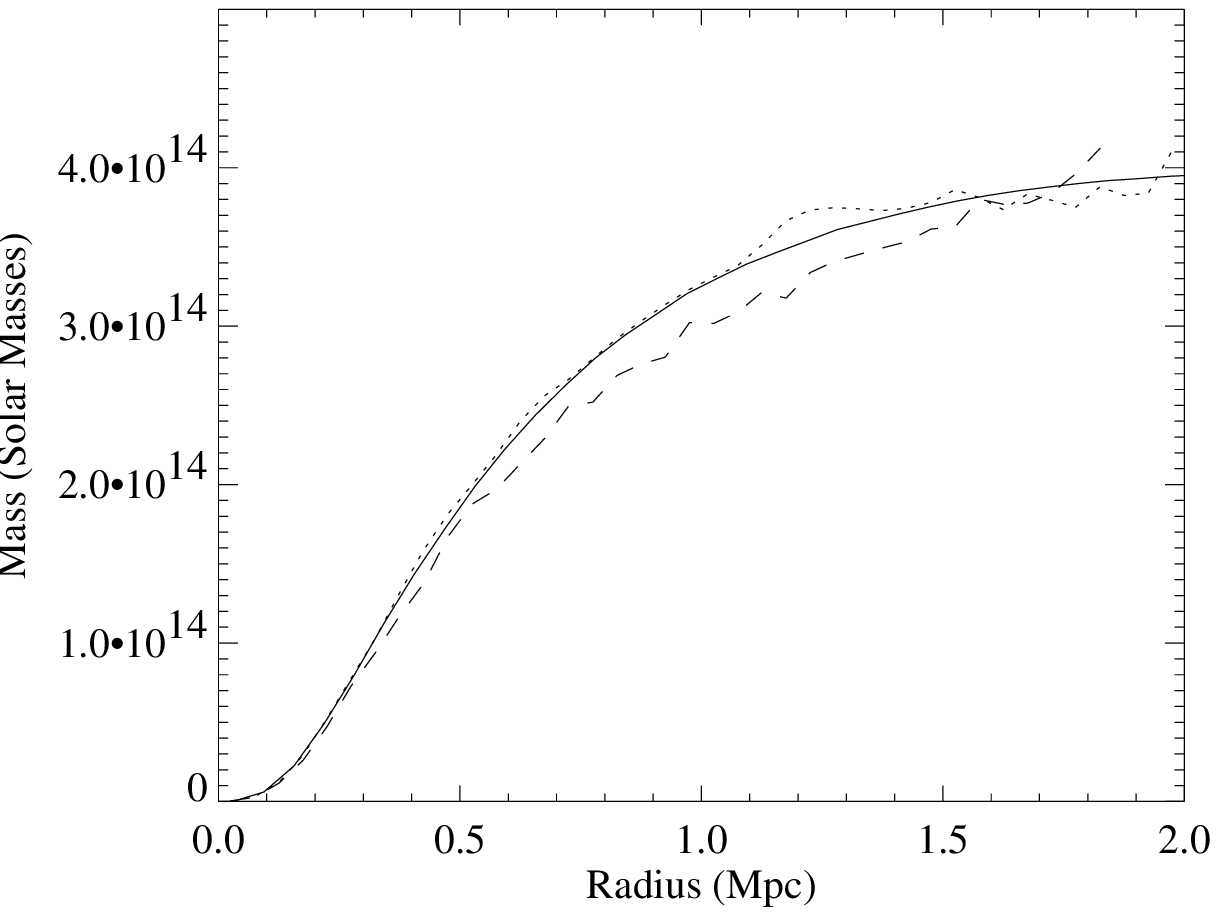}
\caption[Mass Estimates for a Hydrostatic Cluster]
{ }
\label{staticmass}
\end{figure}

\clearpage
\begin{figure}[htbp]
\centering \leavevmode
\epsfxsize=0.85\textwidth \epsfbox{ 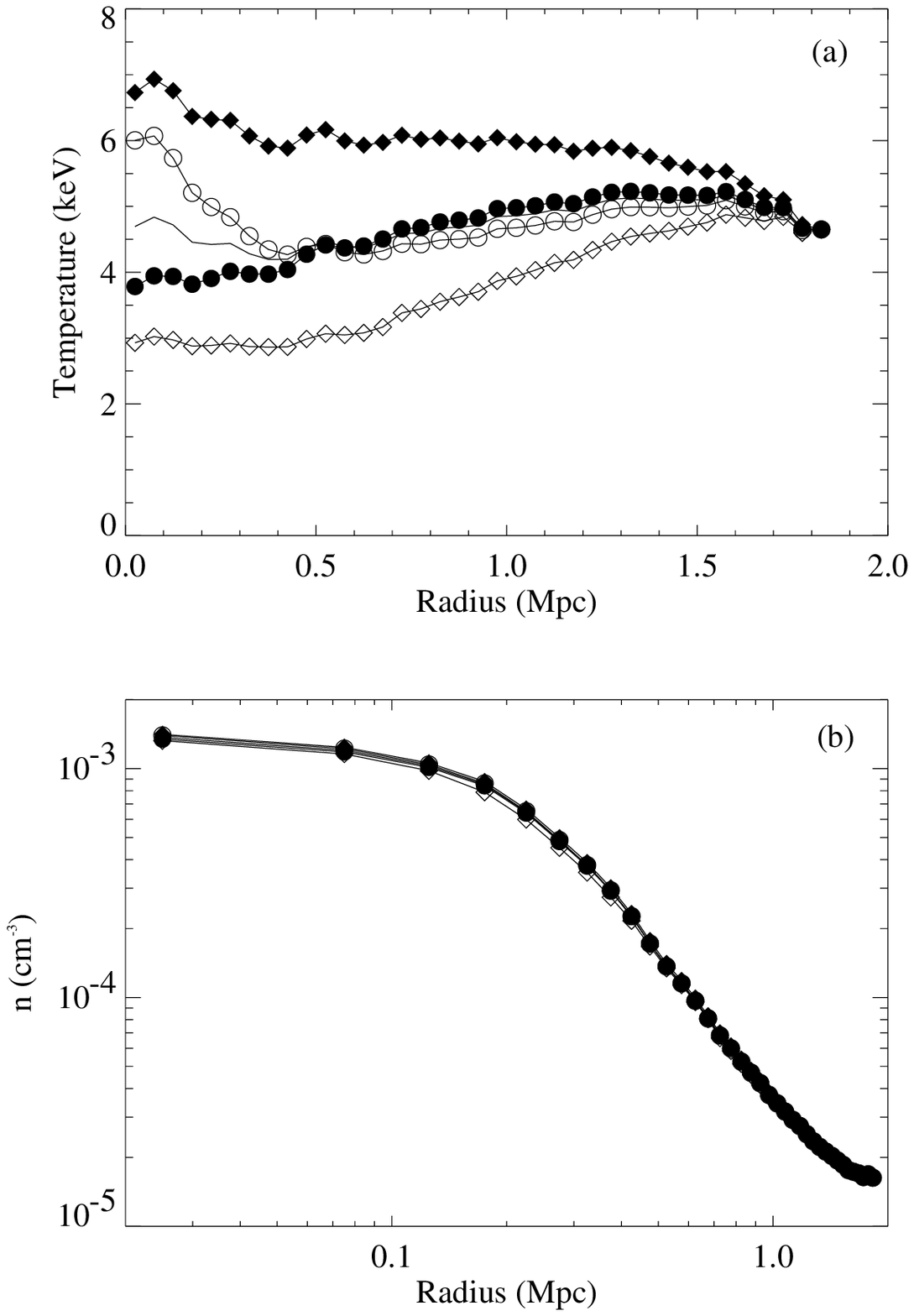}
\caption[Systematic Errors in the X-ray Deprojection Profiles]
{ }
\label{syserr}
\end{figure}

\begin{figure}[htbp]
\centering \leavevmode
\epsfxsize=0.65\textwidth 
\vspace{7.5in}
\includegraphics{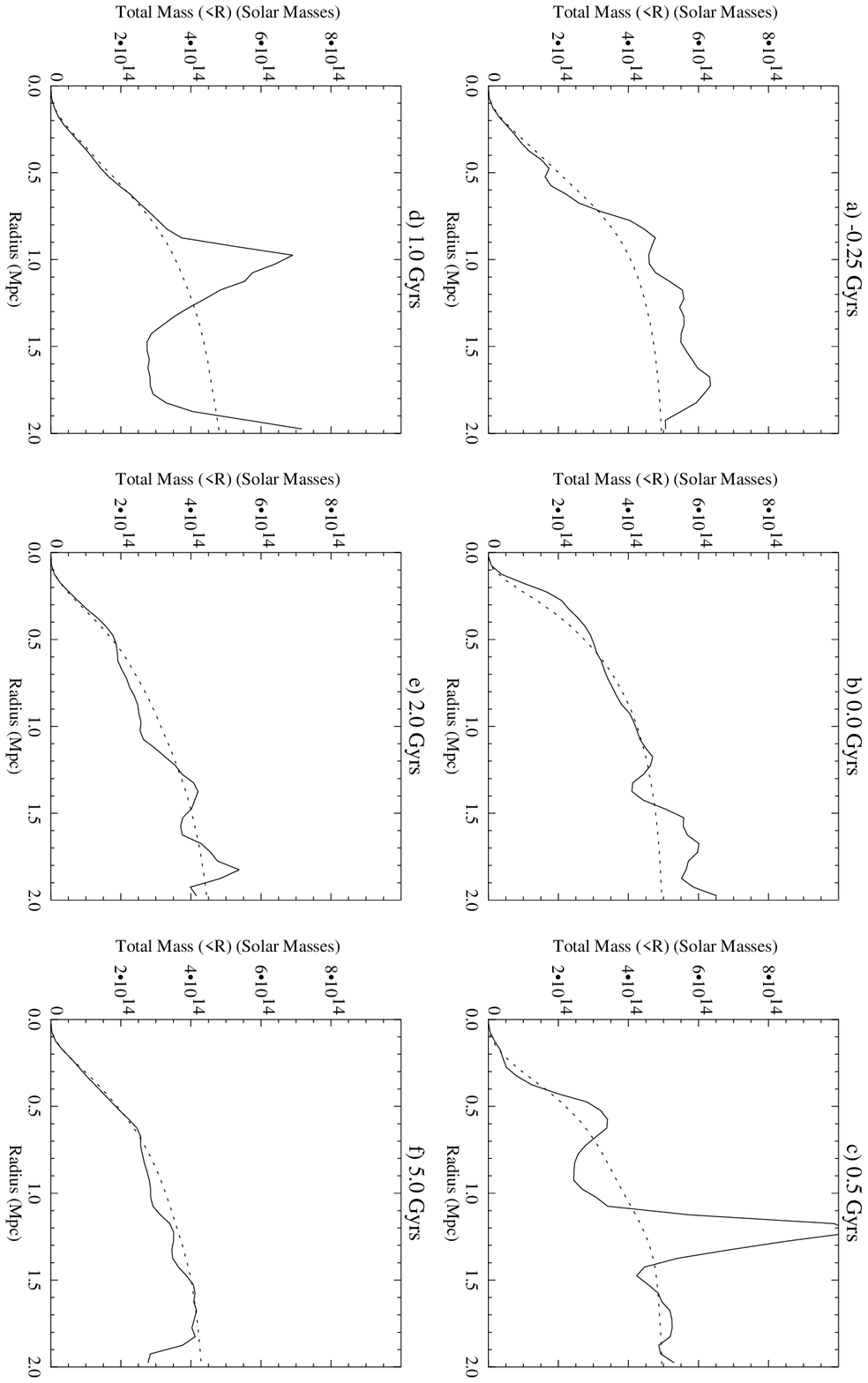}
\caption[Equilibrium-Base Mass Estimates]
{ }
\label{eqmass1}
\end{figure}

\begin{figure}[htbp]
\centering \leavevmode
\epsfxsize=0.7\textwidth \epsfbox{ 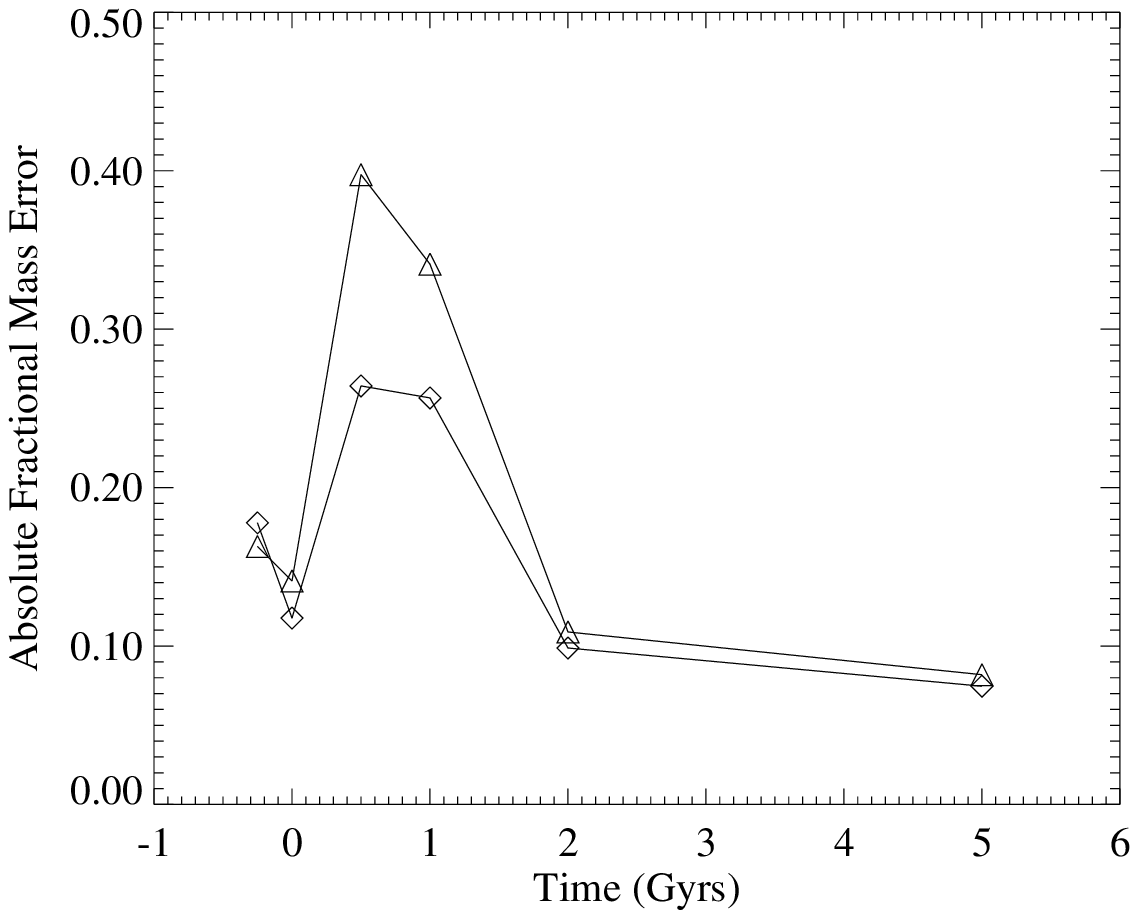}
\caption[Error in the Equilibrium-Base Mass Estimates]
{ }
\label{eqmerr1}
\end{figure}

\begin{figure}[htbp]
\centering \leavevmode
\epsfxsize=0.8\textwidth 
\vspace{7.5in}
\includegraphics{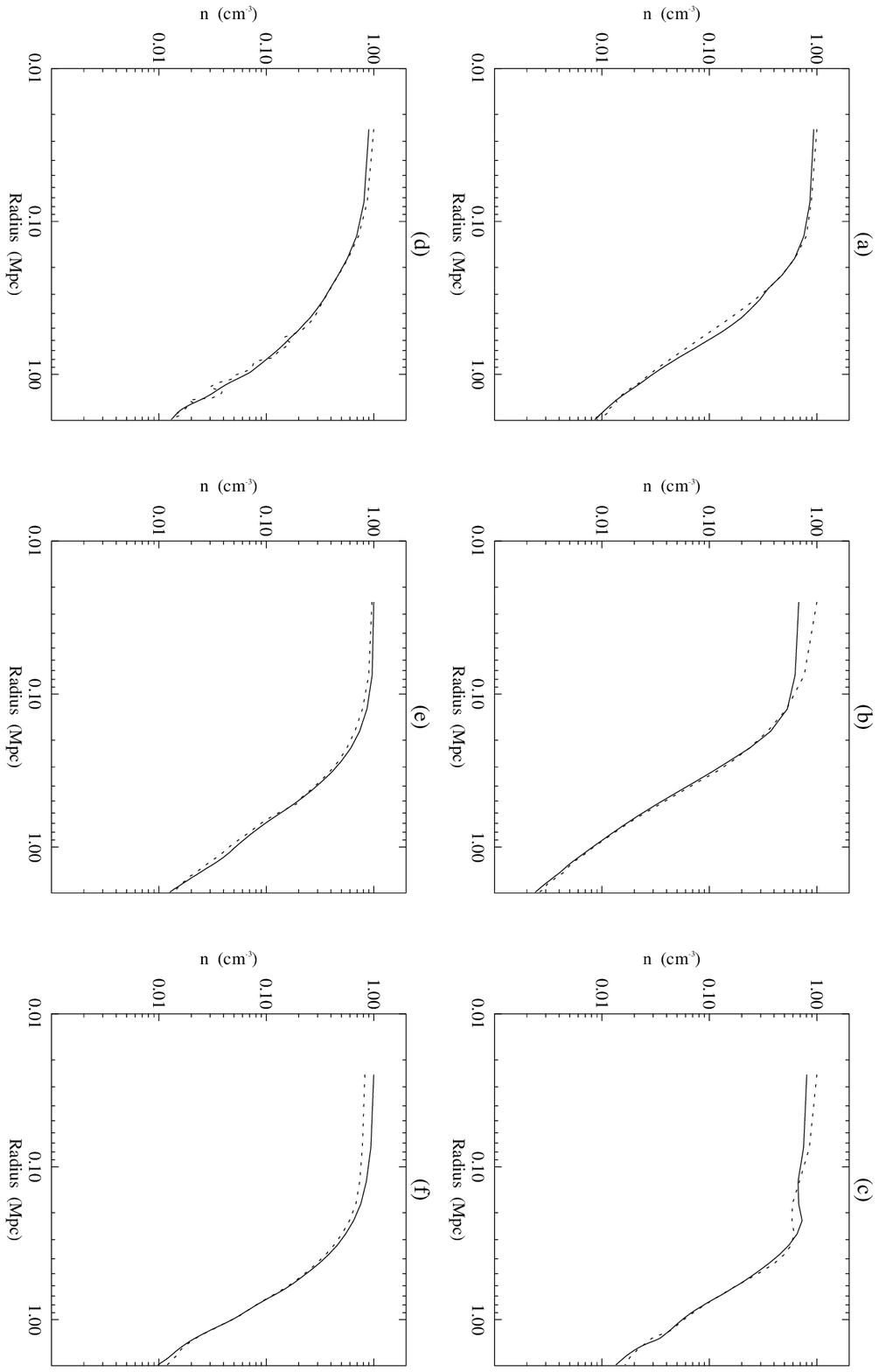}
\caption[X-ray Deprojected Gas Density Profiles]
{ }
\label{depdens}
\end{figure}

\begin{figure}[htbp]
\centering \leavevmode
\epsfxsize=0.8\textwidth 
\vspace{7.5in}
\includegraphics{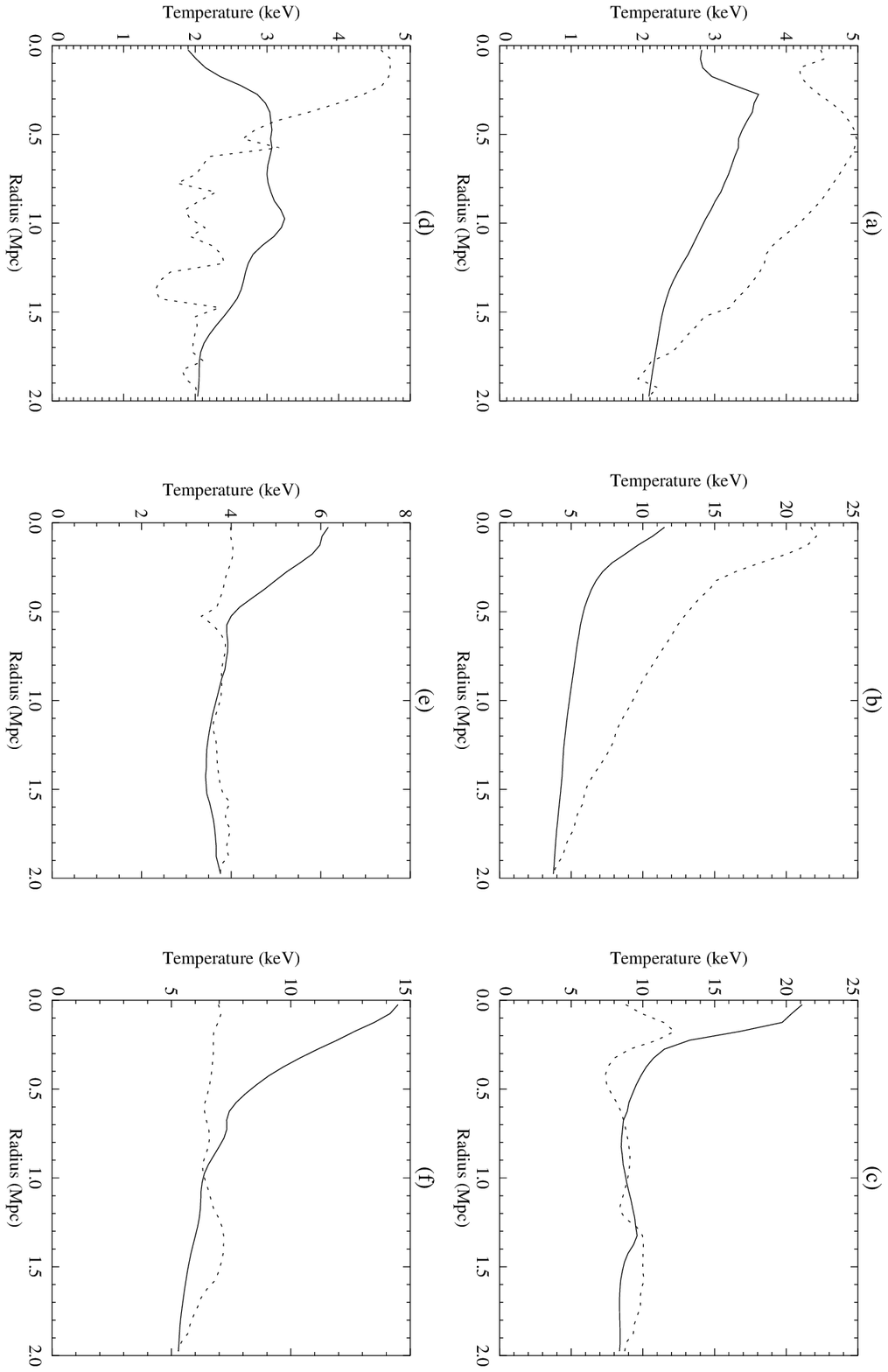}
\caption[X-ray Deprojected Temperature Profiles]
{ }
\label{deptemp}
\end{figure}

\clearpage
\begin{figure}[htbp]
\centering \leavevmode
\epsfxsize=0.8\textwidth 
\vspace{7.5in}
\includegraphics{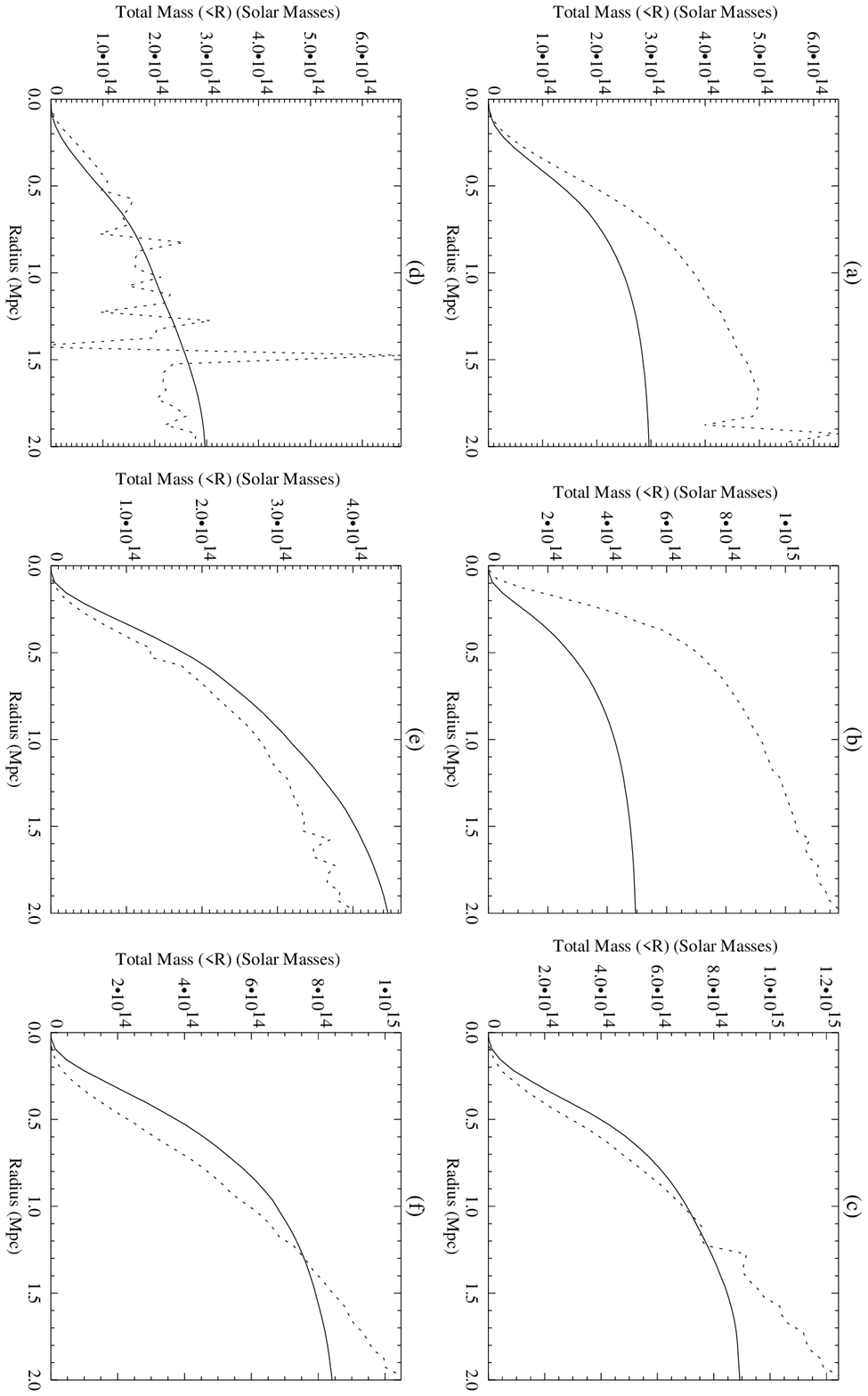}
\caption[X-ray Deprojection-Based Mass Estimates]
{ }
\label{depmass}
\end{figure}

\begin{figure}[htbp]
\centering \leavevmode
\epsfxsize=0.65\textwidth \epsfbox{ 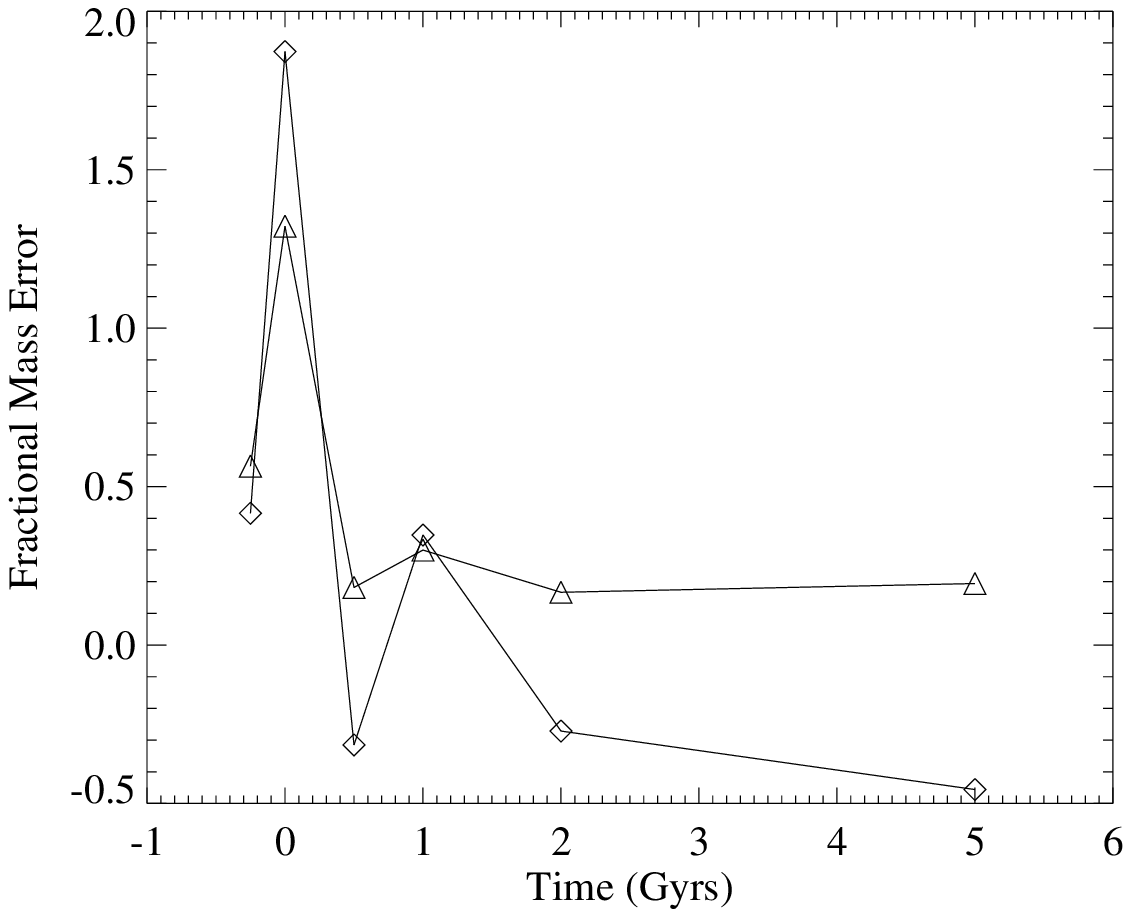}
\caption[Errors Associated with Deprojection-Based Mass Estimates]
{ }
\label{depmerr1}
\end{figure}

\begin{figure}[htbp]
\centering \leavevmode
\epsfxsize=0.65\textwidth \epsfbox{ 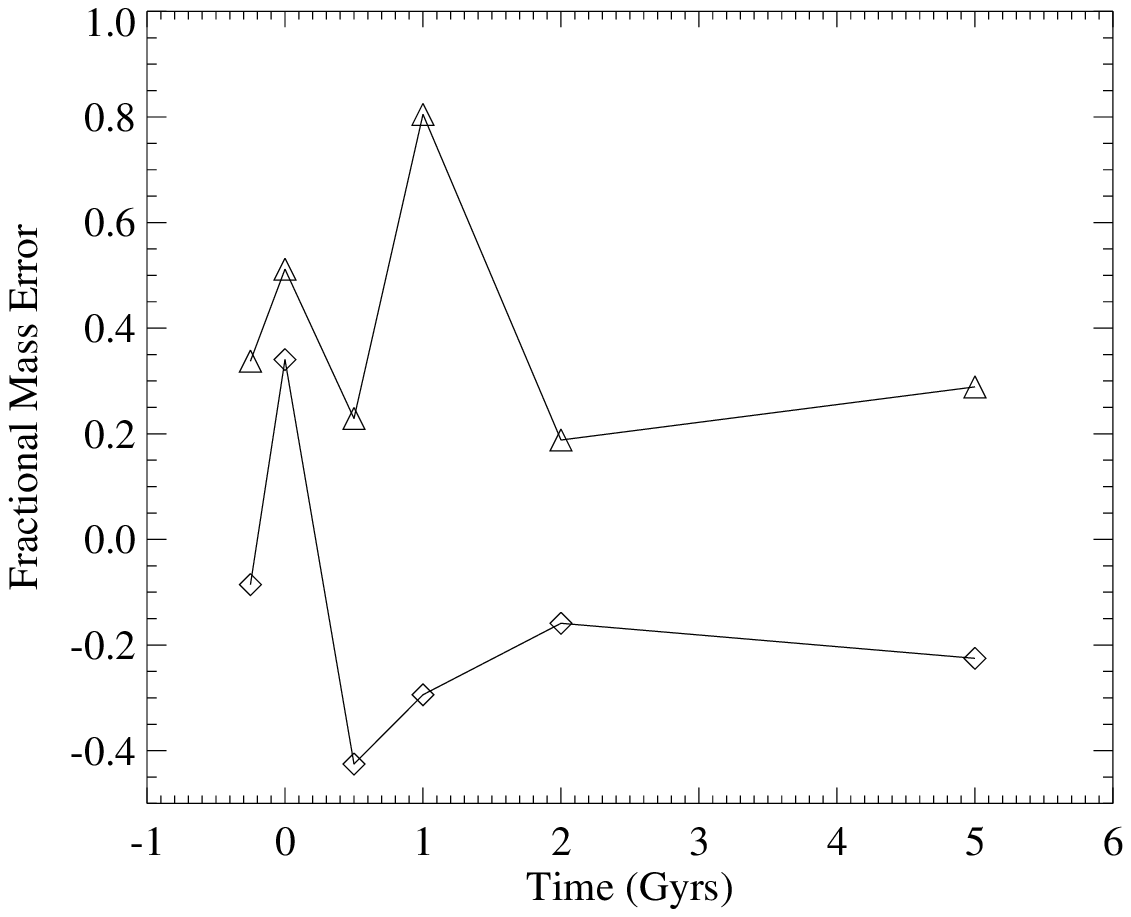}
\caption[Errors Associated with Isothermal Mass Estimates]
{ }
\label{depmerr2}
\end{figure}

\begin{figure}[htbp]
\centering \leavevmode
\epsfxsize=0.75\textwidth 
\vspace{7.0in}
\includegraphics{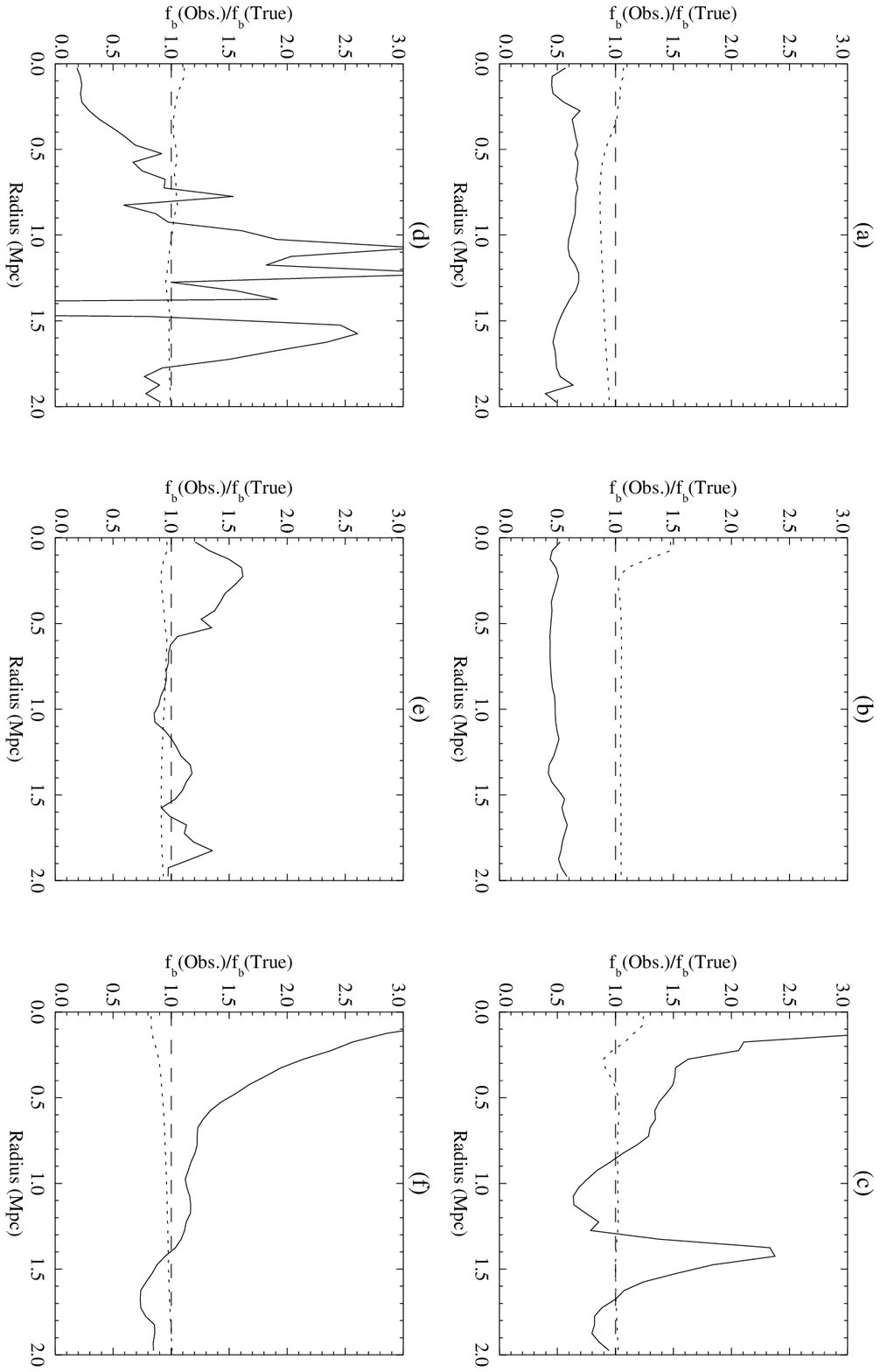}
\caption[Errors in the Cluster Baryon Fraction]
{ }
\label{bafract}
\end{figure}

\end{document}